\documentclass[twocolumn,preprintnumbers,
  amsmath,amssymb,amsfonts, final, a4paper,aps,
  citeautoscript,superscriptaddress,footinbib]{revtex4-1}

  \usepackage{times}
  \usepackage{graphicx}
  \usepackage[latin1]{inputenc}
  \usepackage{mathrsfs}
  \usepackage[vcentermath]{youngtab}
  \usepackage{float}
  \usepackage{color}

  \usepackage{url}

  \usepackage[colorlinks,bookmarks=false,
     citecolor=blue,linkcolor=red,urlcolor=blue]{hyperref}

  \usepackage[normalem]{ulem}
  \usepackage{xspace}

  \def\Jl{J_\Vert}  
  \def\Jp{J_\perp}  
  \def\Jca{\Jp^{c1}}  
  \def\Jcb{\Jp^{c2}}

  \def\jpc{1.562}

  \def\jpa{0.581}
  \def\jpb{0.596}

  \def\SdotS{\langle S \cdot S\rangle}
  \def\SdotSleg{\SdotS _{\Vert}}
  \def\SdotSrung{\SdotS _{\perp}}

  \def\nS{n_\mathrm{S}}
  \def\nSx{\langle \nS \rangle}
  \def\nSxx{\langle \nS(x) \rangle}
  \def\NS{N_\mathrm{S}}
  \def\NSx{\langle \NS \rangle}
  \def\Nm{m}
  \def\nuS{\langle \nu_\mathrm{S} \rangle}

  \def\qinc{q}

  \def\kf{k_{F}}

  \newcommand{\Fig}[1]{Fig.~\ref{#1}}
  
  \newcommand{\FIG}[1]{Figure~\ref{#1}}

  \newcommand{\Sec}[1]{Sec.~\ref{#1}}
  \newcommand{\Eq}[1]{Eq.~\eqref{#1}}
  \newcommand{\App}[1]{App.~\ref{#1}}

  \def\Tr{\ensuremath{\mathrm{Tr}}}
  \newcommand{\trace}[1]{\mathrm{Tr}(#1)}

  \newcommand{\SU}[1]{\ensuremath{\mathrm{SU}(#1)}}

  \DeclareFontFamily{OT1}{pzc}{}
  \DeclareFontShape{OT1}{pzc}{m}{it}{<-> s * [1.200] pzcmi7t}{}
  \DeclareMathAlphabet{\mathpzc}{OT1}{pzc}{m}{it}
  \newcommand{\rsym}{\ensuremath{\mathpzc{r}}\xspace}

  \newcommand{\be}{\begin{equation}}
  \newcommand{\ee}{\end{equation}}
  \newcommand{\bea}{\begin{eqnarray}}
  \newcommand{\eea}{\end{eqnarray}}

  \nocite{White92,Schollwoeck05,Schollwoeck11,Wb12_SUN,Li13,Liu15}

  \def\DMRGrefs{White92,*Schollwoeck05,*Schollwoeck11}
  \def\SUNrefs{Wb12_SUN,*Li13,*Liu15}

\makeatletter

\begin{document}

\title{
  Unified Phase Diagram of
  Antiferromagnetic \SU{N} Spin Ladders
}

\author{A. Weichselbaum} 
\email[e-mail: ]{weichselbaum@bnl.gov}
\affiliation{Physics Department, Arnold Sommerfeld Center
for Theoretical Physics, and Center for NanoScience,
 Ludwig-Maximilians-Universit{\"a}t, 80333 Munich, Germany}
\affiliation{Department of Condensed Matter Physics and Materials
Science, Brookhaven National Laboratory, Upton, NY 11973-5000, USA}

\author{S. Capponi} 
\affiliation{Laboratoire de Physique Th\'eorique, CNRS UMR 5152,
  Universit\'e Paul Sabatier, F-31062 Toulouse, France.}

\author{P. Lecheminant} 
\affiliation{Laboratoire de Physique Th\'eorique et Mod\'elisation,
  CNRS UMR 8089, Universit\'e de Cergy-Pontoise, Site de Saint-Martin,
  F-95300 Cergy-Pontoise Cedex, France}

\author{A. M. Tsvelik} 
\affiliation{Department of Condensed Matter Physics and Materials
Science, Brookhaven National Laboratory, Upton, NY 11973-5000, USA}

\author{A. M. L{\"a}uchli} 
\affiliation{Institut f\"ur Theoretische Physik,
Universit{\"a}t Innsbruck, A-6020 Innsbruck, Austria}

\date{\today}

\begin{abstract}

Motivated by near-term experiments with ultracold
alkaline-earth atoms confined to optical lattices, we
establish numerically and analytically the phase diagram
of two-leg \SU{N} spin ladders. Two-leg ladders provide
a rich and highly non-trivial extension of the single chain
case on the way towards the relatively little explored
two dimensional situation. Focusing on the experimentally
relevant limit of one fermion per site, antiferromagnetic
exchange interactions, and $2\leq N \leq 6$, we show that
the phase diagrams as a function of the interchain (rung)
to intrachain (leg) coupling ratio $\Jp/\Jl$ strongly
differ for even vs. odd $N$.
For even $N=4$ and $6$, we demonstrate that the phase
diagram consists of a single valence bond crystal (VBC) with
a spatial period of $N/2$ rungs.
For odd $N=3$ and $5$, we find surprisingly rich phase
diagrams exhibiting three distinct phases.  For weak rung
coupling, we obtain a VBC with
a spatial period of $N$ rungs, whereas for strong coupling
we obtain a critical phase related to the case of a single
chain.  In addition, we encounter intermediate phases for odd
$N$, albeit of a different nature for $N=3$ as compared to
$N=5$.  For $N=3$, we find a novel gapless intermediate phase
with $\Jp$-dependent incommensurate spatial fluctuations in a
sizeable region of the phase diagram.  For $N=5$, there are
strong indications for a narrow potentially gapped
intermediate phase, whose nature is not entirely clear.
Our results are based on
(i) field theoretical techniques,
(ii) qualitative symmetry considerations, and 
(iii) large-scale density matrix renormalization group (DMRG)
simulations keeping beyond a million of states by fully
exploiting and thus preserving the \SU{N} symmetry.

\end{abstract}

\maketitle

\section{Introduction} \label{sec:intro}

Two-leg spin ladders have been  a focus of much
theoretical and experimental work over more than
two decades.  This strong interest stems in part
from purely theoretical reasons \cite{Dagotto96,
Gogolin98-boso, Giamarchi04-P1D, Schmidinger13}
but is also motivated by experimental realizations
\cite{Azuma94,Caron11, Luo13}. 
These simple magnetic quantum
systems might also be employed as quantum simulators
for fundamental theories of particle and many body
physics \cite{Ward13-qsim, Lake10,Jordens08}.
From a theoretical point of view, ladder systems
provide the simplest non-trivial step away from
well-understood purely one-dimensional systems 
to higher dimensions.
Quite interestingly, a confinement of fractional quantum
number excitations  can be investigated in simple two-leg
spin ladders. In the case of two spin-half \SU{2}
Heisenberg chains coupled by an antiferromagnetic
spin-exchange interaction, the theory predicts that
gapless fractional spin-half excitations of individual
chains (spinons) are confined into gapful spin-1
excitations (triplons) even by an infinitesimal interchain
coupling \cite{Shelton96}.  Experiments provide evidence
for such confinement \cite{Schmidinger13,Lake10}.

In condensed matter systems, the SU(2) symmetry is quite
natural due to the mostly electronic origin of the quantum
magnetism.  In recent years it has, however, been
theoretically proposed \cite{Cazalilla2009,Gorshkov2010}
to create \SU{N} quantum magnets by loading fermionic
alkaline-earth-like atoms into optical lattices and driving
them into the Mott insulating regimes.
Considerable experimental progress by several groups
\cite{Zhang14,Scazza14, Cappellini14, Pagano14, Bloch15}
shows this to be an interesting further avenue for quantum
magnetism.
Experiments can realize two-leg \SU{N} spin ladders
from ultracold alkaline-earth or ytterbium atoms.
The ladder geometry may be realized by double-well
optical lattices similar to those in two-leg Bose-Hubbard
models with cold atoms \cite{SStrabley06, Danshita07,
Atala14}.

When dealing with fermions, a major challenge consists
in cooling them down to reach ground-state properties.
In that context, \SU{N} symmetry is appealing since a larger
number of flavors ($N$) allows to reach lower temperature
(or entropy)~\cite{Hazzard2012}. Therefore short-distance
correlations characteristic of Heisenberg physics should be
accessible, e.g. as  discussed for one-dimensional systems
\cite{Messio12,Bonnes12} or even two-dimensional \SU{N}
Hubbard models \cite{Cai13}.
Recent developments of quantum gas microscopy also
promise increased access to measure local quantities
\cite{Jordens08, Cheuk15, Parsons15,Greif16}
or short-distance correlations \cite{Greif11, Parsons16}.

From the theoretical point of view, the one-dimensional
\SU{N} Heisenberg chain is well characterized thanks
to the exact solution \cite{Sutherland75}, and it
represents a critical state quite analogous to the \SU{2}
case.  When considering more general single-band Hubbard
models of \SU{N} fermions, the phase diagram becomes
already richer with the emergence of different kind of
Mott insulating phases or dimerized phases \cite{Capponi16}. 
Moreover, it appears that additional degrees
of freedom (such as an orbital one) could give rise to even
more exotic phases, such as symmetry-protected topological
ones \cite{Capponi16}. In two dimensions, there has been a
large number of studies devoted to various lattices and $N$.
Restricting to `minimal' models where, at each site, the local
Hilbert space corresponds to the fundamental \SU{N}
representation (i.e.~one particle per site in a fermionic
language), analytical arguments~\cite{Hermele2009,Hermele11}
have triggered a series of numerical investigations which have
proposed, for instance, that the ground-state can break
spontaneously the \SU{N} symmetry~\cite{Toth10,Bauer12} or
break some lattice symmetries~\cite{Corboz13,Nataf16} or break
both symmetries~\cite{Corboz11}, or  exhibit an algebraic
spin-liquid phase~\cite{Corboz12b}. Now, it has to be
mentioned that the two-dimensional simulations remain quite
challenging, so that a deep and detailed understanding of the
two-leg ladder case could shed some light on non-trivial
effects when going from one to two dimensions.  The inherent
difficulty to numerically simulate large $N$ in \SU{N} could
also be a further niche for quantum simulations.  Last but not
least, we will argue that a minimal model is experimentally
feasible since it requires to have \SU{N} symmetry,
one particle per site (which is good to prevent losses),
and ladder geometry, all of which have been realized in
experiments.  

The paper is organized as follows. In the remainder of this
section, we define the model system (subsection~\ref{sec:model}),
and give a brief overview of what is known about the phase
diagrams of \SU{N} ladder, the open questions and a summary of
our resulting phase diagram (subsection~\ref{sec:overview}).
In \Sec{sec:CFT} we present our analytical analysis for weak
rung coupling based on conformal field theory (CFT) and mean
field theory.  In \Sec{sec:DMRG} we present our exact numerical
results for the full range of antiferromagnetic rung coupling
based on the density matrix renormalization group (DMRG). In
\Sec{sec:outlook} we give a summary, followed by an outlook.


\begin{figure}
\includegraphics[width=0.95\linewidth,clip]{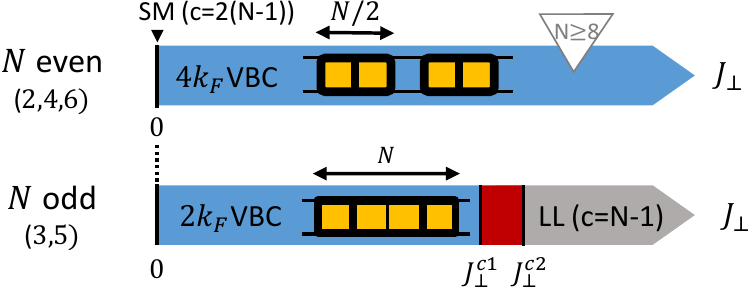}
\caption{
  (Color online) Phase diagrams of two-leg \SU{N}
  antiferromagnetic ladders for $N \leq 6$ vs. $\Jp$,
  setting $\Jl=1$. At $\Jp=0$, all ladders are
  critical [two decoupled Sutherland models (SM), with
  total central charge $c=2\times (N-1)$].
  For finite $\Jp>0$, for even $N$ a single symmetry-broken
  $4\kf$ valence-bond crystal (VBC) phase with an $N$-site
  plaquette is found, which for $N=2$ reduces to a trivial
  nondegenerate rung singlet phase
  (note that for $N\geq 8$ and large
  $J_\perp$, a critical phase with $c=N-1$ is expected
  even for even $N$).
  For odd $N$, in stark contrast, we find three phases.
  For small $\Jp$ a $2\kf$-VBC, for large $\Jp$ a critical
  Luttinger liquid (LL) with central charge $c=N-1$, and
  an intermediate phase. This intermediate phase for $N=3$
  ($\Jp \in [\sim1, \jpc]$) is incommensurate, and for
  $N=5$ ($\Jp \in [\jpa,\jpb]$) is reminiscent of
  topological phases.
}
\label{fig:overview}
\end{figure}


\subsection{Model system}
\label{sec:model}

Motivated by ultracold atomic gases of alkaline-earth-like
fermionic atoms, such as Yb ($N$ up to $6$)  and Sr ($N$
up to $10$), confined to an optical lattice, we start our
description with an $N$-species single-band fermionic
Hubbard model in a two-leg ladder geometry:
\begin{eqnarray}
  {\cal H}_\mathrm{Hubbard}
   &=& 
     \sum_{\ell x} \bigl(
      - \mu\,n_{\ell x}
      + \tfrac{U}{2} n_{\ell x} (n_{\ell x}-1)
     \bigr)   \label{eq:Hubbard} \\
    &-& \sum_{x \alpha} \Bigl(
       t_\perp 
          c^\dagger_{1 x \alpha} c^{\phantom{\dagger}}_{2 x \alpha}
     + t_\parallel \sum_{\ell} 
       c^\dagger_{\ell x \alpha} c^{\phantom{\dagger}}_{\ell,x+1,\alpha}
       + \mathrm{H.c.} \Bigr), \notag
\end{eqnarray}
where $x$ denotes the position along the ladder, $\ell \in
\{1,2\}$ labels the two legs, and $c^\dagger_{\ell x \alpha}$
and $c^{\phantom{\dagger}}_{\ell x \alpha}$ are the fermionic
creation and annihilation operators of a fermionic atom at
lattice position $(x,\ell)$ in the internal state $\alpha$,
and $n_{\ell x} = \sum_{\alpha}c^\dagger_{\ell x \alpha}
c^{\phantom{\dagger}}_{\ell x \alpha}$.  The hopping
amplitudes along the leg (rung) are labeled as $t_\parallel$
($t_\perp$), respectively, and can easily be controlled in
experiments by the strength of the optical lattice.  The
on-site interaction $U>0$ is assumed to be positive,
throughout.
We assume the chemical potential $\mu$ to be chosen
such that we have a filling of one fermion per site
(and therefore  Fermi wave-vector $\kf\equiv\pi/N$)~%
\footnote{In cold atom experiments $\mu$ also possibly
accounts for the (harmonic or box-form) trap.  In the Mott
insulating regime we expect a large central region where
the density is pinned to the desired value~\cite{Bloch15}}.
Then the limit of strong on-site repulsion $U\gg
t_\parallel, t_\perp$ leads to a Mott insulator, where the
charge fluctuations are suppressed, while the only
remaining degrees of freedom are the \SU{N} spins at each
lattice site. 
Their local state space is described by a single multiplet
of dimension $N$ that transforms in the {\it fundamental}
irreducible representation of the \SU{N} symmetry group
which, in the language of Young tableaux, corresponds to
a single box \mbox{( ${\tiny \yng(1)}$ )}.
In second order perturbation theory in the hopping,
virtual charge fluctuations induce an effective dynamics
in the spin-only description, which is called the \SU{N}
Heisenberg model:
\begin{equation}
  {\cal H}_\mathrm{ladder} =  J_{\parallel}
    \sum_{\ell,x,A} 
      S^{A}_{\ell,x}  S^{A}_{\ell,x+1}
  + J_{\perp} \sum_{x,A} S^{A}_{1,x}  S^{A}_{2,x}
\text{ ,}\label{eq:H:spinladder}
\end{equation}
where $S^{A}_{\ell,x}$ with $A=1, \ldots, N^2 -1$ denote the
\SU{N} spin operators i.e.~the generators of the \SU{N} Lie
algebra.
The antiferromagnetic exchange coupling constants
$J_{\parallel}$ and $J_{\perp}$ are related to the parameters
of the underlying Hubbard model as $J_{\parallel(\perp)} =
4\,(t_{\parallel(\perp)})^2/ U$ in the leading order of
perturbation theory.  By using the generalized Pauli identity
(see \App{app:dmrg}),
\begin{equation}
   S_{i} \cdot S_{j} \equiv
   \sum_A S^{A}_{i}  S^{A}_{j} = \tfrac{1}{2} ( P_{i,j}-\tfrac{1}{N} )
\text{ ,}\label{eq:parity}
\end{equation}
with the permutation operator $P_{i,j} = \sum_{\alpha,\beta}
|\alpha_i,\beta_j\rangle\langle \beta_i ,\alpha_j|$
for sites $i\equiv(\ell,x)$, the spin Hamiltonian can
be rewritten as a sum of two-site permutations (up to
constant terms), if the spin operators act on the
fundamental representation, as is the case here.
With $\langle P_{i,j} \rangle \in [-1,1]$, the generalized
``spin-spin'' correlations are constrained to the range
$\langle S_i \cdot S_j \rangle \in [-\tfrac{1}{2}, +\tfrac{1}{2} ]
- \tfrac{1}{2N}$.  Conversely, the symmetric (+) or
antisymmetric (-) weight on the bond between sites $i$ and
$j$ can be simply obtained by
\begin{equation}
  p_{ij}^{\pm} \equiv \langle \tfrac{1}{2} (1\pm
  P_{ij})\rangle = \tfrac{1}{2} \pm \bigl( \langle S_i
  \cdot S_j \rangle + \tfrac{1}{2N}\bigr)
\text{ .}\label{eq:parity:2}
\end{equation}

In this paper we now proceed to establish the phase diagrams
of the spin Hamiltonian ${\cal H}_\mathrm{ladder}$ in
\Eq{eq:H:spinladder} for $N$ varying from $2$ to $6$
analytically and numerically for the full range of $\Jp/\Jl$
in the experimentally accessible antiferromagnetic regime
$\Jl,\Jp \geq 0$.  This ties together partly available
literature for small and large ladder coupling $\Jp$
\cite{Affleck88, Dufour15, Capponi16, Bossche01, Lecheminant15,
James2017}, and the known case of $N=2$ \cite{Shelton96}.
We also speculate on the physics for larger $N$.
While it would be desirable to explore $N$ up to $10$ for
experiments on Sr atoms, for the Heisenberg model as well
as for the Hubbard model for the whole $U>0$ range, these
goals are currently out of reach for exact numerical
approaches.
After all, the models inherit the numerical complexity
of $N$-flavor many-body systems. For \SU{N} symmetric models
this manifests itself predominantly in the fact that typical
individual multiplets grow exponentially in size with
increasing $N$ (in practice, $d\lesssim 10^{N-1}$).

\subsection{Overview of the Phase Diagrams}
\label{sec:overview}

In this paper, we determine the phase diagrams as a
function of an antiferromagnetic interchain exchange
interaction $J_{\perp}>0$ for up to $N=6$. We 
work at zero temperature, fixing $J_\parallel=1$ as the unit
of energy, unless stated otherwise. The goal of this
section is to present the key results of this paper
in a compact way and to refer the reader to the following
sections for a more detailed presentation organized by
the technique.

A schematic summary of our result is presented in
\Fig{fig:overview}. These phase diagrams have been obtained
using field theoretical results detailed in \Sec{sec:CFT}
and large-scale DMRG simulations with results detailed in
\Sec{sec:DMRG}. 

Let us start with some known limits of these phase
diagrams and a statement of the hitherto open questions.
The case $N=2$ corresponds to the well studied \SU{2},
$S{=}1/2$ two-leg spin ladder \cite{Shelton96},
which we include for completeness only. It has been
established that the entire $J_\perp>0$ range forms
one phase which is continuously connected to the
non-degenerate product state of rung singlets which one
obtains in the limit $\Jp \rightarrow +\infty$. Hence the
name "rung singlet phase" encountered in the literature.
This phase has a unique ground state on open and periodic
systems and a gap to all excitations.

A different established limiting case is the strong
coupling limit, i.e. $\Jp \rightarrow +\infty$ for all
$N$,  where the system can be described effectively as
a single \SU{N} Heisenberg chain:
\begin{equation}
{\cal H}_\mathrm{chain} =  J_\mathrm{chain}
    \sum_{x,A} 
      \tilde{S}^{A}_{x}  \tilde{S}^{A}_{x+1}
\text{ ,}\label{eq:spinchain}
\end{equation}
albeit with different local spin degrees of freedom,
in that $\tilde{S}^{A}_{x}$ acts within the two-box
antisymmetric representation ( ${\tiny \yng(1,1)}$ )
of dimension $N(N-1)/2$. The exchange constant
$J_\mathrm{chain}$ is proportional to $J_\parallel$ in
general. While this is a trivial Hamiltonian and state
for $N=2$ as discussed in the paragraph before, the
situation is much richer for larger $N$ where we 
have to distinguish based on the parity of $N$
\cite{Affleck88,Dufour15}: for even $N=4$ or $N=6$,
the chain will spontaneously dimerize or trimerize,
respectively, which in the original ladder language
corresponds to $2 \times 2 = 4$-site and $3 \times 2
= 6$-site plaquette formation (see \Fig{fig:overview})
\cite{Capponi16, Dufour15}. For the \SU{4} case, the
plaquette formation has already been reported for a
ladder when $\Jp$ is large \cite{Bossche01}.
For odd $N>1$ and even $N>6$ (see App.~D of
\cite{Lecheminant15} and Ref.~\cite{Dufour15}, and also
\Fig{fig:overview}), the chain remains critical in the
\SU{N}$_1$ Wess-Zumino-Novikov-Witten (WZNW) universality
class, which corresponds to a highly symmetric
$(N-1)$-component Luttinger liquid (LL).

Finally the last (partially) known starting point is the
weak coupling limit. For $J_\perp=0$, the system decouples
into two copies of \SU{N} Heisenberg spin chains, one on
each leg, with the spins in the fundamental representation.
The individual spin chain models are also known as \SU{N}
Sutherland models~\cite{Sutherland75}.  The Sutherland model
is integrable by means of Bethe ansatz and displays quantum
critical behavior in the \SU{N}$_1$ WZNW universality class
with central charge $c=N-1$~\cite{Sutherland75,Affleck88}.
Recent field theoretical work by two of the present
authors~\cite{Lecheminant15} has established that for all
{\em even} $N$ an infinitesimal interchain coupling
$J_\perp>0$ is expected to give a gapped phase of the
valence bond crystal (VBC) type. A cartoon description of
these states is given by an array of \SU{N} singlets
consisting of $N$ spins on $N/2$ consecutive rungs
(see \Fig{fig:overview}).
For even $N>2$ this therefore amounts to a state with broken
translation symmetry along the ladders and an ensuing ground
state degeneracy of $N/2$ in the thermodynamic limit. The
system is expected to exhibit gaps for magnetic and
non-magnetic excitations beyond the ground state manifold.
These types of VBCs with a unit cell of $N$ spins in a
$N/2\times 2$ geometry may also be called a ``$4\kf$ VBC''
since their modulation wave vector $2\pi/(N/2)=4\pi/N$
corresponds to four times $\kf$, having $\kf=\pi/N$.
The {\em odd} $N$ case is also unstable upon switching
on a finite $J_\perp$. However the nature of the resulting
phases were not clear so far.  In this paper we show
analytically that for $N=3$ the system forms a different
VBC type as compared to the even $N$ case.  Here a
``$2\kf$~VBC'' forms with a unit cell comprising $2N$
spins arranged in a $ N\times 2$ building block along the
ladder (see \Fig{fig:overview}).
This picture is confirmed by large scale DMRG simulations.
Our numerical results show that the $2\kf$ VBC phase also
persists for $N=5$ at weak coupling. 

Aside from this review of known limiting cases, the
physics in the entire intermediate coupling regime for all
$N>2$ is basically unexplored. From a qualitative point of
view it is possible that the physics for {\em even}
$N\leq6$ is adiabatically connected from weak to strong
coupling, although this has not been demonstrated before.
However the phases at weak and strong coupling for {\em
odd} $N=3,5$ manifestly differ, and thus require at least
one phase transition in between.

In the following we demonstrate that the even cases $N=4$
and $N=6$ are indeed forming a single $4\kf$-VBC phase for
all $\Jp>0$, while the odd cases $N=3$ and $N=5$ exhibit
extended intermediate phases between the $2\kf$-VBC phases
at small $\Jp$ and the critical regime at strong rung
coupling.  The intermediate phase for $N=3$ is of a novel
\SU{N} incommensurate type, where the $\Jp$-dependent
incommensuration is driven by the relative population
density of the symmetric \mbox{( ${\tiny \yng(2)}$ )} and
antisymmetric \mbox{( ${\tiny \yng(1,1)}$ )} rung multiplets.
In the case $N=5$ the intermediate phase is quite narrow,
potentially gapped, and generally of rather different
nature as compared to the $N=3$ intermediate phase.

\section{Weak rung coupling results $\Jp\gtrsim 0$: \newline
Field theoretical approach} \label{sec:CFT}

In this section we start the construction of the phase
diagrams coming from the decoupled chain limit
$J_\perp=0$. Using field theoretical methods we establish
the fate of this singular starting point after switching
on a small antiferromagnetic rung coupling $J_\perp$.
Previously established results for all even $N$ are
reviewed. Then the case $N=3$ is treated explicitly.

\subsection{Continuum description}

As stated above, at $J_{\perp} = 0$, model
(\ref{eq:H:spinladder}) becomes two decoupled \SU{N}
Sutherland models with a total central charge $c=2(N-1)$. The
low-energy properties of the Sutherland model can be obtained
by starting from the U($N$) Hubbard model at $1/N$ filling
with a large repulsive $U$ interaction \cite{Affleck86,
Affleck88, Assaraf99,Manmana11}. At low energy below the
charge gap, the \SU{N} operators in the continuum limit are
described by \cite{Affleck86,Affleck88}:
\begin{equation}
   S^{A}_{\ell,x} \simeq J^{A}_{\ell L} (x) +  J^{A}_{\ell R} (x)
   + e^{ i 2k_F x} i \lambda   \; \Tr\bigl( g_\ell (x) T^A \bigr)
   + \mathrm{H.c.}
\text{ ,}\label{eq:spinop}
\end{equation}
where $J^{A}_{\ell L,R}$ are the left and right $\SU{N}_1$
currents, and $T^A$ are generators of the su($N$) algebra in
the fundamental representation (Gell-Mann matrices) normalized
such that $\Tr( T^A T^B ) = \delta^{AB}/2$.  The $2k_F$ term
involves the $\SU{N}_1$ WZNW field $g_\ell$ with scaling
dimension $\frac{N-1}{N}$.  The non-universal real constant
$\lambda$ corresponds to the average over the charge degrees
of freedom in the large $U$ limit.

The continuum limit of the two-leg \SU{N} spin ladder
(\ref{eq:H:spinladder}) is then described by the
low-energy Hamiltonian:
\begin{eqnarray}
  {\cal H} &=& {\cal H}^{(1)}_{\SU{N}_1}
             + {\cal H}^{(2)}_{\SU{N}_1} + \lambda_1 \int dx \;
   \bigl[ \Tr(g_1 g_2^+) + \mathrm{H.c.} \bigr] \nonumber \\
   &+& \lambda_2 \int dx \;
   \bigl[ \Tr g_1 \Tr g_2^{+}  + \mathrm{H.c.} \bigr]
\text{ ,}\label{eq:cftgen}
\end{eqnarray}
where ${\cal H}^{(\ell)}_{\SU{N}_1}$  denotes the Hamiltonian
of the $\SU{N}_1$ CFT for the $\ell=1,2$ decoupled chain,
and $\lambda_1 = J_{\perp} \lambda^2 /2$, $\lambda_2 =
- \lambda_1/N$. Model (\ref{eq:cftgen}) describes two
$\SU{N}_1$ WZNW models perturbed by two strongly relevant
perturbations with the same scaling dimension $2(N-1)/N <
2$.  Since these perturbations are strongly relevant, we
have not included in \Eq{eq:cftgen} the marginal
current-current interaction.  The effective field theory
(\ref{eq:cftgen}) can be simplified by exploiting the fact
that when $J_{\perp} \ne 0$ the continuous symmetry group
of  model (\ref{eq:H:spinladder}) is $\SU{N} \times
{\mathbb{Z}}_2$ making it more natural to consider the
following conformal embedding \cite{Francesco97-CFT}:
\begin{equation}
   \SU{N}_1 \times \SU{N}_1 \sim \SU{N}_2 \times \mathbb{Z}_N
\text{ .}\label{conformalembedding}
\end{equation}
Here the $\SU{N}_2$ CFT has central charge $c=2
\tfrac{N^2-1}{N+2}$ and ${\mathbb{Z}}_N$ is the parafermionic
CFT with central charge $c = \frac{2(N-1)}{N+2}$. It describes
universal properties of the phase transition of the
${\mathbb{Z}}_N$ generalization of the 2D Ising model
\cite{Zamolodchikov85}.
The low-energy Hamiltonian (\ref{eq:cftgen}) can then be
expressed in the new basis and reads \cite{Lecheminant15}
\begin{eqnarray}
   {\cal H} &=& {\cal H}_{\SU{N}_2} + \bigl[
      {\cal H}_{{\mathbb Z}_N} - g \int d x \;  \left( \Psi_{1L}  \Psi_{1R}
    + \mathrm{H.c.} \right) \bigr]  \nonumber
\\
   &+& \lambda_2 \int dx   \; \Tr(\Phi_{\rm adj})
   \bigl( \sigma_2 + \sigma_2^{\dagger} \bigr)
\text{ ,}\label{cftnewbasis}
\end{eqnarray}
where $g = -N J_{\perp} \lambda^2 /8\pi^2$ and $\lambda_2 =
-J_{\perp} \lambda^2 /2N$. The effective field theory
(\ref{cftnewbasis}) contains two different sectors, the \SU{N}
singlet one in square brackets, and the magnetic one, which
depends on the \SU{N} degrees of freedom. These sectors are
coupled by the last term with coupling constant $\lambda_2$.
Operators ${\cal H}_{\SU{N}_2}$ and $\Phi_{\rm adj}$ act in
the \SU{N} sector; the former one is the Hamiltonian of the
$\SU{N}_2$ CFT and the latter one is a primary field with
scaling dimension $\Delta_{\rm adj} = \tfrac{2N}{N+2}$ which
transforms in the adjoint representation of the \SU{N}.
In \Eq{cftnewbasis}, ${\cal H}_{{\mathbb Z}_N}$ is the
Hamiltonian of ${\mathbb{Z}}_N$ CFT which is generated
by the ${\mathbb{Z}}_N$ parafermion currents $\Psi_{1L,R}$
with conformal weights $h, {\bar h} = \tfrac{N-1}{N}$.
Operators $\sigma_k$ ($k=1, \ldots N-1$) are the
${\mathbb{Z}}_N$ order parameter fields with scaling
dimensions $ \Delta_{k} = \tfrac{k(N-k)}{N(N+2)}$.
Fields with $k >1$ are generated by fusion of the
fundamental order parameter field $\sigma_1$; $\sigma_k$
take non-zero expectation values in the  phase which
spontaneously breaks the ${\mathbb{Z}}_N$ symmetry. 

\subsection{Weak-coupling phase diagram} 

The two perturbations in \Eq{cftnewbasis} are relevant
with the same scaling dimension $\tfrac{2(N-1)}{N} < 2$.
The crucial difference between \SU{2} and \SU{N>2} cases
stems from two  facts. First, for $N=2$, $\sigma_2$ is
just an identity operator and hence  the two sectors of
the theory remain decoupled ~\cite{Shelton96}. Second, for
$N=2$, $\Psi_{1L,R}$ are Majorana fermions and $\Phi_{\rm
adj}$ is also bilinear in Majorana fermions. In this
respect, model (\ref{cftnewbasis})  can be expressed in
terms of four massive Majorana fermions which account for
the emergence of non-denegerate gapful phases for any sign
of $J_{\perp}$~\cite{Shelton96}. The situation is much
more involved in the case of $N>2$ due to the coupling of
the magnetic and the singlet sectors. Hence a different
physical picture is expected.  One way to determine the
low-energy properties of model (\ref{cftnewbasis}) is to
use the fact that when $\lambda_2 =0$, the ${\mathbb{Z}}_N$
sector of (\ref{cftnewbasis}) is an integrable deformation
of the ${\mathbb{Z}}_N$ parafermions introduced by
Fateev~\cite{Fateev91,Fateev91-Z}. Since the infrared
properties of this theory  strongly depend on the parity
of $N$ we have to consider odd and even cases separately.

\subsubsection{Even $N$}

For even $N$, a spectral gap is generated for the
${\mathbb{Z}}_N$ Fateev model (the expression in the
square brackets in (\ref{cftnewbasis})) for any sign of
$J_{\perp}$. In Ref. \onlinecite{Lecheminant15}, two of us
have exploited the exact solution of the Fateev model to
find the low-energy properties of weakly-coupled \SU{2n}
two-leg spin ladder. In the simplest $N=4$ case, we have
found the competition between a tetramerized ($2 \kf$)
phase with a four-fold ground state degeneracy and
a two-fold degenerate plaquette (dimerized, $4 \kf$)
phase~\cite{Lecheminant15}. In particular, when
$|\lambda_1| \ll | \lambda_2|$, a plaquette phase was
predicted. At sufficiently strong $\Jp\gg 1$, we already
know that there is spontaneous translation symmetry
breaking with the formation of VBC formed with 4-site and
6-site plaquettes respectively for $N=4$ and $N=6$
respectively. However, at weak coupling the field-theory
analysis of Ref.~\onlinecite{Lecheminant15},  has not made
it clear which phase wins the competition when both
$\lambda_1$ and $\lambda_2$ are similar in magnitude as it
is the case when the  interchain coupling $J_{\perp}$ is
switched on. Hence it may happen that the latter phase is
replaced by a tetramerized phase ($2\kf$ VBC). To settle
this issue we resort to a numerical investigation in Sec.~\ref{sec:dmrg_N_even}.

\subsubsection{Odd $N$}

The strategy of Ref.~\cite{Lecheminant15} is not
applicable for odd $N$ since the physical properties of
the integrable Fateev ${\mathbb{Z}}_{N}$ model strongly
depend on the sign of its coupling constant $\lambda_1$
\cite{Fateev91,Fateev91-Z}. In particular, when
$J_{\perp}>0$, the ${\mathbb{Z}}_N$ Fateev model
(\ref{cftnewbasis}) with $\lambda_2 = 0$ displays
an integrable massless renormalization group (RG) flow from the ${\mathbb{Z}}_N$
ultraviolet (UV) fixed point to the infrared (IR) one governed
by the minimal model ${\cal M}_{N+1}$ CFT with central
charge $c = 1 - \tfrac{6}{(N+2)(N+1)}$ \cite{Fateev91-Z}.
In the simplest case, i.e. $N=3$, one has a massless flow
from ${\mathbb{Z}}_3$ CFT to the ${\cal M}_{4}$ CFT which
corresponds to the $c=7/10$ tricritical Ising model (TIM)
CFT  which describes the critical properties of
two-dimensional discrete models with ${\mathbb{Z}}_2$
Ising spins $\sigma = \pm 1$ and a vacancy variable
$t=0,1$ \cite{Francesco97-CFT}.  The strategy for odd $N$
is thus to rewrite model (\ref{cftnewbasis}) in the IR
limit by exploiting the existence of the massless RG flow.
To this end, we need the UV-IR transmutation of the
${\mathbb{Z}}_{N}$  fields and most importantly the
expression of the field $ \sigma_2 + \sigma_2^{\dagger}$
at the IR ${\cal M}_{N+1}$ fixed point. Unfortunately,
to the best of our knowledge, we are aware of such UV-IR
transmutation only in the simplest $N=3$ case. In the odd $N$ case, we can thus only investigate the low-energy properties of the SU(3) spin ladder in this paper. In this respect, the transmutation of the ${\mathbb{Z}}_{3}$ primary
fields has been investigated in Ref.~\onlinecite{Klassen93}
by means of the thermodynamic Bethe ansatz
\cite{Zamolodchikov91, Klassen92}.   The most important one for our study is the UV-IR
transmutation of the ${\mathbb{Z}}_3$ spin field 
\begin{eqnarray}
\sigma =
\sigma_2^{\dagger}, ~~ \sigma +  \sigma^{+}  \rightarrow
\epsilon_{\rm TIM}, \label{UVIRZ3bis}
\end{eqnarray}
where $\epsilon_{\rm TIM}$ is the
so-called thermal operator of TIM CFT, which has  scaling
dimension $1/5$~\cite{Francesco97-CFT}.  We thus deduce
the IR limit of model (\ref{cftnewbasis}) in terms of the
fields of the TIM $\times\ \SU{3}_2$ CFT:
\begin{eqnarray}
   {\cal H} =  {\cal H}_{\SU{3}_2}  + {\cal H}_{{\rm TIM}}    
   + \kappa \int dx   \; \Tr(\Phi_{\rm adj}) \; \epsilon_{\rm TIM},
\label{cftnewbasisIR}
\end{eqnarray}
where ${\cal H}_{{\rm TIM}}$ is the Hamiltonian of the TIM
CFT and $\kappa \simeq \lambda_2 <0$ when $J_{\perp} >0$.
The interacting part of model (\ref{cftnewbasisIR}) is a
strongly relevant perturbation with scaling dimension $7/5
< 2$ which should open a mass gap for both sectors of the
theory. To investigate the IR properties of model
(\ref{cftnewbasisIR}) we resort to a simple mean-field
approach to decouple the $\SU{3}_2$ and TIM sectors and to
exploit the existence of a massive integrable perturbation
of the TIM CFT: ${\cal H}_{\rm mf} = {\cal H}_1 + {\cal
H}_2$ with:
\begin{eqnarray}
   {\cal H}_1 &=& {\cal H}_{{\rm TIM}} + \kappa \int dx \;
    \langle \Tr(\Phi_{\rm adj})  \rangle \; \epsilon_{\rm TIM} \nonumber
\\
   {\cal H}_2 &=& {\cal H}_{\SU{3}_2} + \kappa \int dx \;
    \langle \epsilon_{\rm TIM}  \rangle \; \Tr(\Phi_{\rm adj})
\label{meanfield}
\end{eqnarray}
The excitation spectra of (\ref{meanfield}) are known.
${\cal H}_1$ describes the TIM CFT perturbed by its
thermal operator which is a massive integrable field
theory for all sign of $\kappa$ \cite{Laessig91}. When
$\kappa < 0$, we have $ \langle \epsilon_{\rm TIM}
\rangle > 0$ which means, within our conventions, that the
underlying 2D TIM belongs to its high-temperature
paramagnetic phase where the ${\mathbb{Z}}_2$ symmetry is
unbroken.  The Hamiltonian  ${\cal H}_2$ of \Eq{meanfield}
describes an $\SU{3}_2$ CFT perturbed by its adjoint
primary field which has been recently studied in Ref.
\onlinecite{Lecheminant15mless}.  A massive behavior
occurs when the coupling constant is negative as in
the $\kappa < 0$ case here. The nature of the physical
properties of the fully gapped phase can be deduced from
the identity:
\begin{eqnarray}
   \Tr(\Phi_{\rm adj}) &=& \Tr( G^{+} T^{A} G T^{A} ) \notag \\
   &\sim&  \bigl( \Tr \; G \;  \Tr \;  G^{+}
  - \tfrac{1}{3} \Tr (G G^{+}) \bigr)
\textbf{ ,}\label{TrPhiadj}
\end{eqnarray}
where $G$ is the WZNW \SU{3} matrix field. We thus find 
that the minimization of model ${\cal H}_2$
with a negative coupling constant selects  the center
group of \SU{3}:
\begin{equation}
   G = \exp( 2 i\pi k/3) I ,
\label{min1adj}
\end{equation}
with $k = 0, 1, 2$ and the solution breaks spontaneously
the one-step translation symmetry $T_{a_0}$. The ground
state may describe thus a trimerized phase. An order
parameter of the latter phase is:
\begin{eqnarray}
   \langle {\cal O}_{+} (n)\rangle &=& e^{- \frac{2 i \pi n}{3}}
   \sum_{\ell=1}^{2}\langle S^{A}_{\ell,n}  S^{A}_{\ell, n+1} \rangle
   \sim \langle  \Tr g_1  + \Tr g_2  \rangle \nonumber
\\
   &\sim& \langle  \Tr G  \; (\sigma +  \sigma^{+} ) \rangle
\text{ .}\label{latticebondphase3}
\end{eqnarray} 
Using  the UV-IR transmutation (\ref{UVIRZ3bis})
we deduce then in the mean-field approximation:
\begin{eqnarray}
   \langle {\cal O}_{+} (n)\rangle 
   \sim \langle  \Tr G  \rangle  \langle  \epsilon_{\rm TIM}\rangle  \ne 0,
\label{latticebondphase3fin}
\end{eqnarray} 
in the ground state (\ref{min1adj}). Let us now consider
the following relative trimerized order operators to have
a physical interpretation of the ${\mathbb{Z}}_2$ symmetry
of the  underlying TIM model:
\begin{eqnarray}
   \langle {\cal O}_{-} (n)\rangle
   &=& e^{- \frac{2 i \pi n}{3}} \langle
     S^{A}_{1,n}  S^{A}_{1, n+1} - S^{A}_{2,n}  S^{A}_{2, n+1} \rangle
     \nonumber \\
   &\sim& \langle \Tr g_1  - \Tr g_2  \rangle
    \sim  \Tr G \; (\sigma - \sigma^{+}) \nonumber \\
   &\sim& \langle \Tr G  \rangle  \langle  \sigma_{\rm TIM} \rangle
\text{ ,}\label{latticebond12}
\end{eqnarray} 
where we have used the the UV-IR transmutation of the
operator $ - i \left(\sigma - \sigma^{+} \right)
\rightarrow \sigma_{\rm TIM}$ along the massless flow
${\mathbb{Z}}_{3} \rightarrow {\rm TIM}$. We thus observe
that the ${\mathbb{Z}}_2$  symmetry of the TIM model
coincides with the permutation $1 \leftrightarrow 2$
symmetry of the two-leg spin ladder. Since the TIM model
belongs to its paramagnetic phase, we have  $\langle {\cal
O}_{-} (n)\rangle = 0$. We thus expect the formation of a
trimerized phase with a three-fold degenerate ground state
where the  \SU{3} singlets are in phase between the
chains.

To summarize, the field theoretical approach predicts
that the decoupled $J_\perp=0$ fixed point is unstable for
all $N$. In the case of even $N$ a small antiferromagnetic
$J_\perp>0$ leads to a valence bond crystal with a unit
cell comprising $N$ sites in a $N/2 \times 2$ geometry,
i.e. an extent of $N/2$ along the ladder. In the language
of a weak-coupling Hubbard model, this corresponds to a
structure with a wave vector $4\kf$.  For $N=3$ the
low-energy approach leads to different valence bond
crystal, here the unit cell consists of $2N$ spin arranged
in a $N\times 2$ geometry with a wave vector $2\kf$.

\section{Numerical results: large scale \newline
\SU{N} DMRG simulations} \label{sec:DMRG}

We simulate the \SU{N} ladders discussed above
initially using exact diagonalization (ED) and conventional
DMRG \cite{\DMRGrefs}, followed by non-abelian DMRG
simulations that are able to fully exploit the
underlying \SU{N} symmetries based on the QSpace
tensor library \cite{\SUNrefs}.
By switching from a state-space based description to a
multiplet-based description, this allows us to keep beyond
$D>500,000$ states for \SU{4} and clearly beyond a million of
states for \SU{5}, corresponding to $D^\ast \leq 4096$
multiplets in either case. A more detailed description of the
numerical aspects is presented in \App{app:dmrg}.

\begin{figure}[!t]
\includegraphics[width=\linewidth]{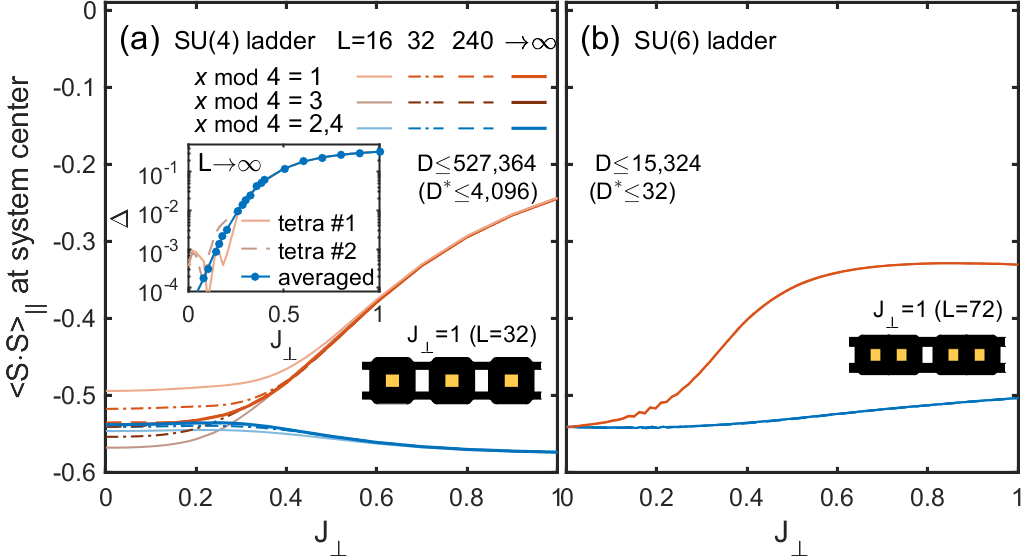}
\caption{(Color online) %
   \SU{N} DMRG results for the nearest-neighbor
   spin correlations $\SdotSleg$ 
   for fixed bonds along the legs as a function of $\Jp$
   for $N=4$ (left panel) and $N=6$ (right panel). In
   both cases we witness the appearance of two different
   sets of values due to the spontaneous
   formation of $2\times 2$ plaquettes for $N=4$ and
   $3\times 2$ plaquettes for $N=6$, as depicted
   in the lower right insets based on actual bond
   energies.
   Panel (a) shows dimerization (and finite-size tetramerization
   for $\Jp \lesssim 0.5$) in uniform \SU{4} ladders
   of length $L=16$, $32$, $240$, together with an
   extrapolation $L\to\infty$ (thick lines) using open
   boundaries. By grouping data at fixed bond position
   $x \mod{N}$, two branches emerge for larger $\Jp$. 
   Tetramerization manifests itself by the additional 
   splitting of branches towards smaller $\Jp$.
   It vanishes in the thermodynamic limit.
   The inset shows the dimerization strength $\Delta$ of
   the extrapolated data ($L\to\infty$), where `tetra \#$n$'
   refers to the dimerization relative to the individual
   tetramerized branches $n=1,2$.
   The analysis for the \SU{6} ladder in panel (b)
   is analogous to panel (a), yet takes data from
   a single DMRG scan [cf.~\App{app:dmrg}] that is
   linear in $\Jp$ over the range shown, using $L=300$.
   By grouping data w.r.t. $x\mod{6}$, again two branches
   emerge, but here due to trimerization (see inset for data
   on an isotropic ladder for $\Jp=1$) with no indication for
   hexamerization.
}
\label{fig:SU46}
\end{figure}

\subsection{Even $N$ \label{sec:dmrg_N_even}}

Our DMRG data for $N=4$ and $N=6$ is presented in
\Fig{fig:SU46}. It demonstrates the realization of a $4\kf =
4\pi/N$ valence bond crystal with a $N/2\times 2$ unit
cell for the entire range $\Jp>0$. Exemplary bond strength
data along uniform ladders is shown in the insets. 
For $N=4$, by inversion symmetry, the rung energies are
uniform throughout the ladder, i.e. are insensitive to the
dimerization. Hence the analysis focuses on the bond
energies along the legs for even $N$, throughout.

For $N=4$ [\Fig{fig:SU46}(a)], careful finite size scaling
shows that the onset of strong tetramerization for
$\Jp\lesssim 0.5$ seen for shorter ladders and suggested
as an actual possible physical phase~\cite{Lecheminant15},
does reduce to a dimerized phase in the thermodynamic limit.
The dimerization strength $\Delta$, defined as the difference
in consecutive $\SdotS_\Vert^\mathrm{NN}$ correlations along
the legs, vanishes only in the limit $\Jp\to 0^+$.
As demonstrated in the center inset of \Fig{fig:SU46}(a),
by averaging data over former tetramerized branches, $\Delta$
diminishes down to below $10^{-4}$ for $\Jp <0.05$.

Similarly, our data for the \SU{6} ladder in \Fig{fig:SU46}(b)
also supports a single symmetry broken
VBC phase. Here a trimerized phase emerges in the entire
antiferromagnetic regime of positive $\Jp$ (see inset).
Given the exponential increase in multiplet size with
increasing $N$ [\onlinecite{\SUNrefs}], however, a full
scale DMRG simulation for \SU{6} remained too costly.
Therefore the data for smaller $\Jp$ needs to be interpreted
more cautiously.  Nevertheless, for larger $\Jp$ converged
data can already be obtained with modest effort ($D^\ast>160$
multiplets, corresponding to $D>87,000$ states, as
used for the uniform ladder in the inset to the panel).

The cases $N=4$ and $N=6$ are thus rather simple in terms of
their phase diagrams, with a single VBC phase spanning the
entire $\Jp>0$ range, analogous to the well studied $N=2$ case
where the nondegenerate rung singlet phase extends over the
same parameter range. It would be interesting, but very
challenging, to study even $N\ge8$, where, as discussed
previously, the large $\Jp$ limit is expected to be \SU{N}$_1$
WZWN critical, while the weak coupling approach predicts a VBC
for small $\Jp$.

\subsection{Odd $N$}

We start with instructive data from exact diagonalization
for $N=3$ shown in \Fig{fig:SU3_Panel-ED}, followed by
large-scale \SU{N} DMRG data for $N=3$ and $N=5$ in
Figs.~\ref{fig:SU3_DMRG} and \ref{fig:SU5_DMRG},
respectively.  In the weak coupling limit, both systems
demonstrate the realization of a $2\kf = 2\pi/N$ valence
bond crystal, as shown in the respective insets of
\Fig{fig:SU3_DMRG}(b) and \Fig{fig:SU5_DMRG}(a)
based on actual bond strength data.
Roughly, this may be interpreted as resonating \SU{N} singlets
formed within $N\times 2$ blocks.  Interestingly, for $N=3$,
this VBC is consistent with the view as an extreme case of a
ladder as a narrow honeycomb lattice strip where such
resonating \SU{3} singlets have been found on 6-site
hexagons~\cite{Corboz13}.

Both systems, $N=3$ as well as $N=5$ also have a critical
phase for $\Jp>\Jcb$ of central charge $c=N-1$ (see
\App{app:largeJp} for excellent numerical
confirmation).  In between the large and small $\Jp$ regime,
however, we find an intermediate phase with highly non-trivial
properties. These two intermediate phases differ drastically
in character for $N=3$ as compared to $N=5$. We investigate
the nature of the intermediate phases in more detail in the
following subsections.

\subsubsection{Intermediate Phase for $N=3$}\label{sec:SU3_incomm}

From previous considerations, we understand the different
phases occurring at weak $\Jp$ coupling [a $N\times 2$ unit
cell VBC] and at strong $\Jp$ coupling [a gapless $SU(N)_1$
WZWN regime]. An earlier numerical and field theoretical
study~\cite{Corboz07}, focussing on a $J_1{-}J_2$ \SU{3}
chain, found a direct continuous transition between a
$SU(3)_1$ WZWN regime and a trimerized VBC phase. One is led
to wonder whether this scenario is also realized here in the
ladder context, since the translation symmetry breaking
aspects of the chain and the ladder VBCs seem related.

A closer inspection, however, reveals that this is {\em
not} the case for our ladder system, and the exchange
symmetry between the two legs of the ladder is responsible
for the discrepancy.  In the large $\Jp$ limit every rung
is in the antisymmetric two-box \SU{N} irreducible
representation ( ${\tiny \yng(1,1)}$ ). But these rung
states are simultaneously also eigenstates of the spatial
leg exchange symmetry ($\ell=1 \leftrightarrow 2$)
with an eigenvalue $-1$.
On the other hand if one inspects the leg exchange symmetry
property of an isolated open-boundary $N\times 2$ ladder
block with $N$ odd at small $\Jp>0$ (as a VBC cartoon
state), one obtains an eigenvalue $+1$, i.e.~a leg-exchange
{\it symmetric} ground state.  In an even further simplified
representation~\footnote{This cartoon representation can be
justified in the fully frustrated model, where an additional
diagonal ladder coupling of equal strength as the leg coupling
renders the local rung spin state a good quantum number, see
Ref.~\protect{\cite{Honecker2000}} for the SU(2) case.} we can
depict an $N=3$ or $N=5$ VBC unit cell as a `product' state
with a [ASA] ($N=3$) or [AASAA] ($N=5$) rung parity pattern,
where A (S) stands for an antisymmetric (symmetric) rung
multiplet. In this cartoon picture, the total leg exchange
eigenvalue is given as the product over all individual
rung parity eigenvalues, and hence changes sign as $\Jp$
is lowered starting from large values.  Note that this
cartoon picture is, in fact, also reflected to some degree
in the actual DMRG data for intermediate $\Jp<\Jca$, e.g.
see rung bond strengths in lower left inset of
\Fig{fig:SU5_DMRG}(a).

The advocated change in the spatial symmetry properties as a
function of $\Jp$ is a strong indication that the symmetric
rung multiplets  have to be taken into account in the ensuing
discussion, while the single chain scenario mentioned above
would be restricted to the sector where all rung multiplets
remain antisymmetric ([AAA] / [AAAAA]). 

It is instructive to rewrite the spin ladder
Hamiltonian~\eqref{eq:H:spinladder} in the rung eigenbasis,
in analogy to the bond operator formulation developed
originally in Ref.~\cite{Sachdev1990} for the \SU{2} case.
In this basis, the rung part of the Hamiltonian is diagonal
and can be brought into a suggestive form, 
\bea
     J_{\perp} \sum_{x,A} S^{A}_{1,x}  S^{A}_{2,x}
 &=& J_{\perp} \underset{\equiv \NS}{\underbrace{\sum_{x} \nS(x)}}
 + \mathrm{const. \ ,}
\eea
with $\nS$ denoting the projector on the symmetric rung
multiplet [cf. \Eq{eq:parity:2}], and therefore $\nSxx$ the
probability, and hence the density of a symmetric rung
multiplet on the rung at position $x$.  Consequently, the
operator $\NS$ counts the total number of rungs in the
symmetric multiplet.  This particular form thus translates the
$\Jp$ coupling into an effective ``chemical potential"
$\mu_S \equiv -J_\perp$ 
for the symmetric rung multiplets with the hardcore
constraint of $\langle \nS \rangle \leq 1$ symmetric
multiplets per rung.

The part of the Hamiltonian involving the leg couplings $\Jl$
is more complicated and we refrain from providing the complete
expression here.  Note, however, that the total Hamiltonian
conserves the parity of the number of symmetric rung states.
This implies that $\langle (-1)^{\NS} \rangle$ is quantized to
$\pm1$ but $\langle \NS \rangle$ itself is not quantized.
Still, we can approximately interpret our numerical results
as though not only the parity of $\NS$ is conserved, but even
$\NS$ itself. As discussed further below, we can indeed count the
number $m$ of changes of $\langle (-1)^{\NS} \rangle$ in the
ground state as a function of $\Jp$ (starting at large $\Jp$)
and loosely relate $\langle \NS \rangle \sim m $. Hence the
symmetric states on rungs behave like effective particles.

We now see a picture emerging where the large $\Jp$ limit
is characterized by the absence of symmetric rung multiplets
in the ground state wave function, because the chemical
potential for the symmetric rung states is exceedingly
large and negative.  Upon lowering $\Jp$, i.e.  increasing
$\mu_S$, it might happen that at one point the lowest
excitation in the other parity sector crosses the current
ground state sector.  Since this changes the total rung
parity, this can lead to various possible phase transition
scenarios, for example a Lifshitz transition where the
ground state is populated by more and more symmetric rung
states at the expense of antisymmetric rung eigenstates,
leading to a novel \SU{N} liquid phase with
$\Jp$-dependent incommensurate correlations.
Another scenario would be a direct first order transition
between the large $\Jp$ critical regime and the VBC at weak
coupling. The VBC can actually also be seen as a "charge
density wave" (CDW) of symmetric rung multiplets with a
finite density $1/N$ of the rungs in the symmetric state.
The question of the nature of the phase diagram in the
intermediate coupling range is thus rephrased into the
question of how the symmetric rung multiplets start to
populate the ladder and ultimately form a period-$N$ CDW
yielding the $2\kf$ VBC. The complementary picture starts
with the CDW/VBC and asks how defects in the form of
reducing $\NS$ achieve to melt the CDW/VBC.


\begin{figure}[!t]
\includegraphics[width=0.95\linewidth]{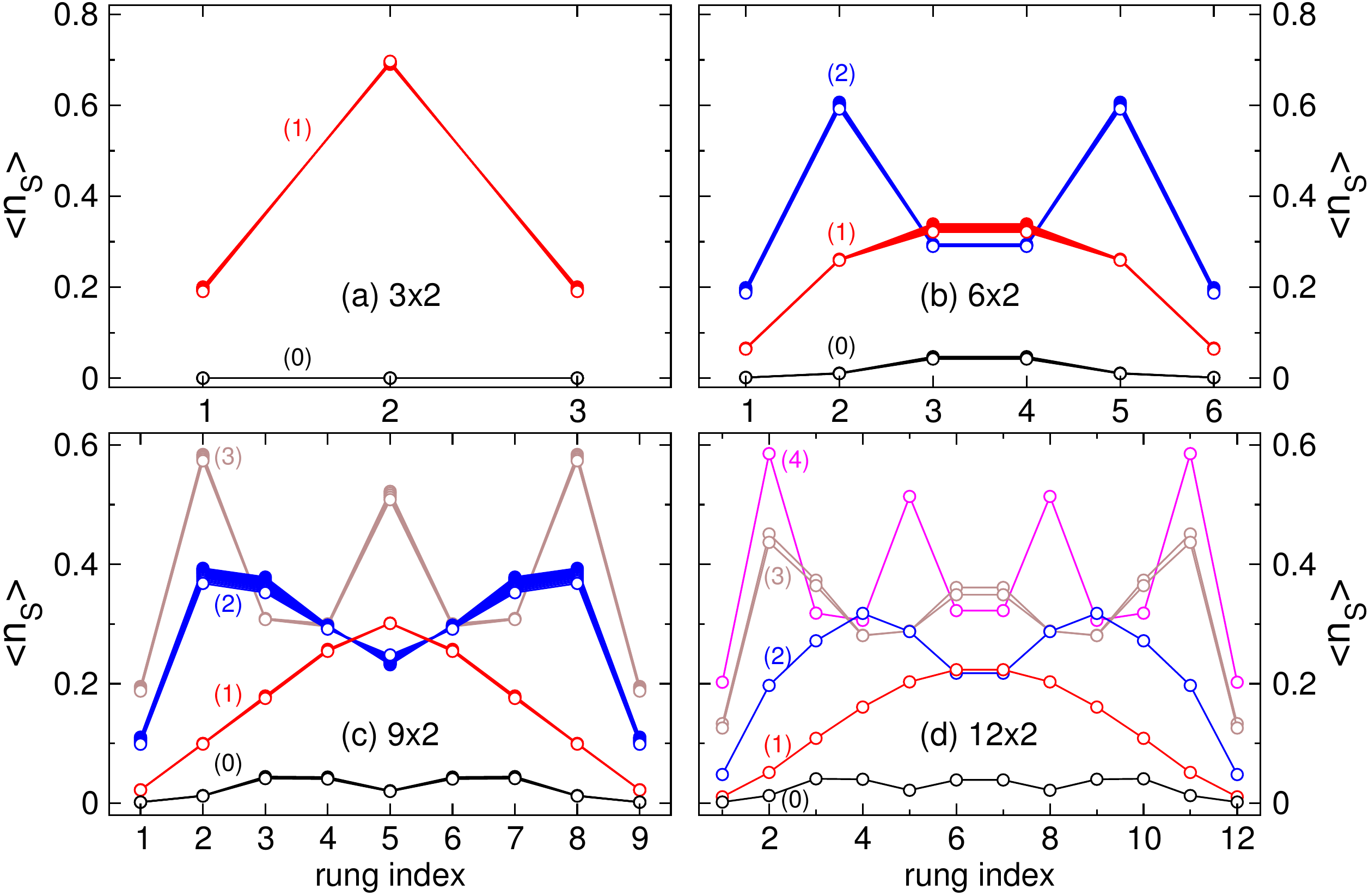}
\caption{(Color online) \SU{3} Heisenberg ladder ---
   ED results showing the local rung parity for small ladders
   with OBC. $\Jp$ varies from $1.35$ to $1.60$ in steps of
   $0.01$ for panels (a) to (c) and in steps of 0.05 for panel
   (d).  The color code groups the $\Jp$ values with similar
   profiles $\nSxx$ [the $\Jp$ values where these profiles switch
   are summarized in \Fig{fig:SU3_ED_Crossings}].
   The number $(\Nm)$ indicates the
   number of significant local maxima, and coincides with the
   number of ground state level crossings encountered, starting
   with $m=0$ when coming from large $\Jp$ (see main text).
} 
\label{fig:SU3_Panel-ED}
\end{figure}

\begin{figure}[!t]
\includegraphics[width=1\linewidth]{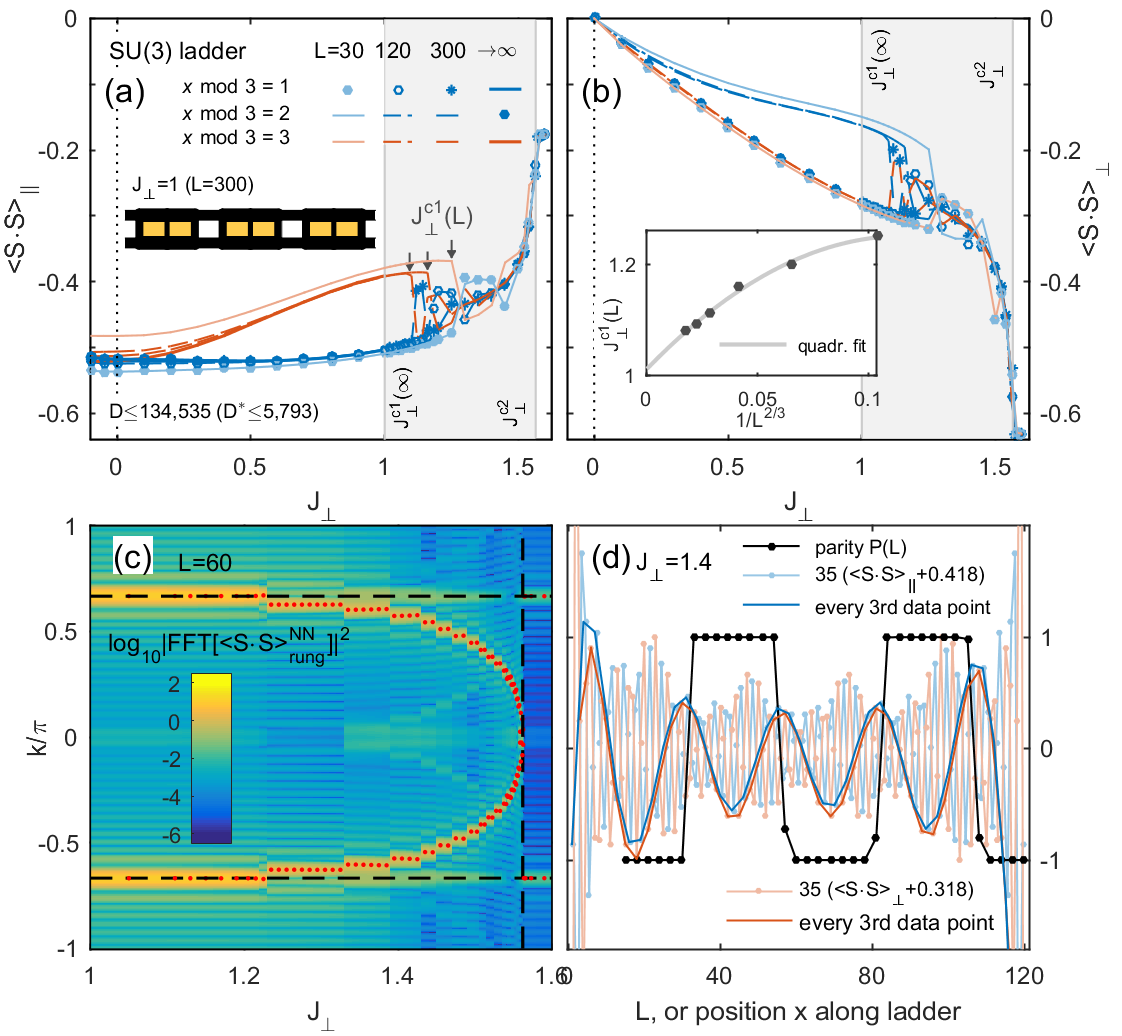}
\caption{(Color online)
  \SU{3} Heisenberg ladder ---
  Panel (a) Bond energies $\SdotSleg$ vs. $\Jp$ along the legs
  in the center of uniform systems (the analysis is completely
  analogous to \Fig{fig:SU46} except that data is grouped
  w.r.t. $x \mod 3$; see legend).  Within the incommensurate
  phase $\Jp \in [ {\sim1}, \jpc]$ (shaded area) the bond
  energies vs. $\Jp$ become irregular.  Inset visualizes the
  bond energies in the center of a uniform $L=300$ ladder at
  $\Jp=1$ (in the VBC phase).
  Panel (b) Bond strengths $\SdotSrung$ along the rungs
  [identical analysis as in (a) otherwise].  Inset shows a
  polynomial extrapolation of $\Jca(L)$ in the thermodynamic
  limit, suggesting $\Jca(L\to\infty)\sim1$.
  Panel (c) Norm-squared (arb. units) of the Fourier transform
  of $\SdotSrung$ along uniforms ladders vs. $\Jp$. The peaks
  related to the incommensurate wave vector $q(\Jp)$
  are tracked by red dots [see also \Fig{fig:SU3_incommL}].
  Additional maxima appear at $k=\pm 2\pi/3$ (horizontal
  black dashed lines) that indicate the superimposed
  3-rung periodicity. The vertical dashed line marks
  $\Jcb$.
  Panel (d) Ground state parity $P(L)$ vs. ladder length
  $L=3p$ with integer $p$, superimposed with scaled
  nearest-neighbor $\SdotSleg(x)$ 
  data along the legs (blue) and rungs (orange),
  all at fixed $\Jp=1.4$. For further details, see
  also \Fig{fig:SU3_incommX}.
} 
\label{fig:SU3_DMRG}
\end{figure}


In order to corroborate these qualitative arguments we resort
to an exact diagonalization study of open boundary $L \times
2$ ladders for a range of $\Jp$ values between $1.35$ and
$1.6$.  In \Fig{fig:SU3_Panel-ED} we display the local rung
density $\langle \nS \rangle$ in the symmetric state as a
function of the rung location $x$ for various $\Jp$ values and
$L=3,6,9,12$ (different subpanels). The considered $\Jp$
values group into sets of curves with similar $\nSxx$ profiles.
The different sets are colored according to the number $(\Nm)$
of significant local maxima we detect.  We use this number
as a proxy for the quantity $\NS$ since, indeed, we find $\NSx
\sim \Nm$.  In the $L=3$ case we either find a completely
vanishing curve [black, $(\Nm=0)$], or a curve with a strong
peak on the central rung [red curve, $(\Nm=1)$].
The remaining panels present the $\nSxx$ profiles for $L=6$,
$L=9$ and $L=12$, providing further evidence for the buildup
of an extended intermediate phase where the $\Jp$ value
controls the number of symmetric rung states $\Nm$ and
therefore the dominant spatial oscillation frequency of rung
bond energies.  In \Fig{fig:SU3_ED_Crossings} in
\App{app:ed_details} we display the $\Jp$ values where
the number $(\Nm)$ of maxima in the $\nSxx$ profile changes by
one. These locations precisely correspond to the anticipated
ground state level crossings, where the total leg-exchange
parity eigenvalue alternates between $+1 \leftrightarrow -1$.

\begin{figure*}[!t]
\centerline{\includegraphics[width=0.9\linewidth]{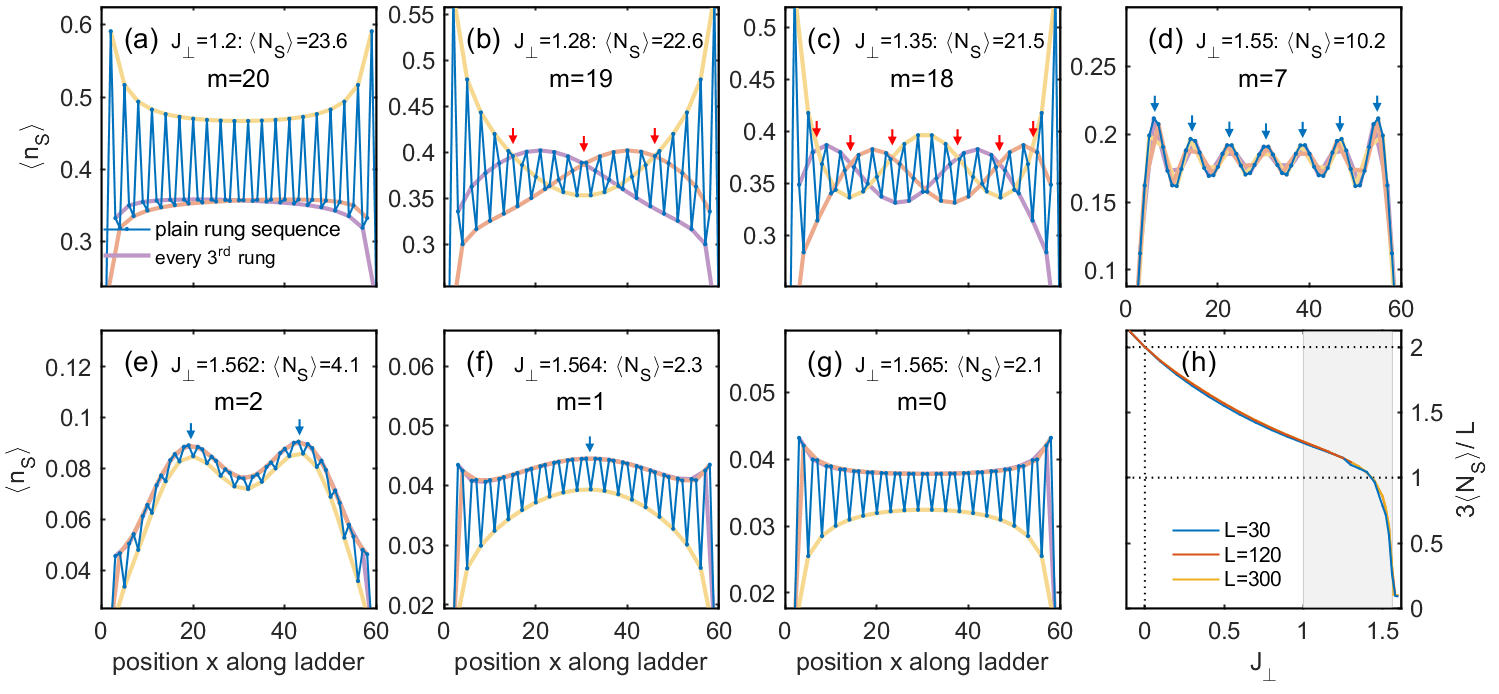}}
\caption{(Color online)
Incommensurate profiles in the \SU{3} Heisenberg ladder
at length $L=60$, using $\nSx = \SdotSrung + \tfrac{2}{3}$
---
   Panels (a-g) show different snapshots across the
   incommensurate regime, with panel (a) already in the
   commensurate small-$\Jp$ regime [note that $\Jca(L=60)
   \gtrsim 1.20$; cf.~\Fig{fig:SU3_DMRG}], whereas panel
   (g) is just outside $\Jcb$.  Small red arrows indicate
   ``domain walls'' (upper crossings amongst light-colored
   lines) which is useful for $\Jp$ closer to $\Jca$.
   Conversely, small blue arrows mark local maxima in the
   plain data sequence (blue) which is useful for larger $\Jp$.
   Every panel also specifies the integrated density
   $\NSx$ of symmetric rung multiplets.
   Panel (h) summarizes $\NSx$ vs. $\Jp$ over the full
   incommensurate range (shaded area) down to $\Jp=0$.
}
\label{fig:SU3_incommD}
\end{figure*}

With these small system exact diagonalizations in mind, we now
switch to a large-scale DMRG study of the physics from weak
to strong $\Jp$ coupling, as summarized in \Fig{fig:SU3_DMRG}.
We first report the $\Jp$ dependence of the nearest-neighbor
spin-spin correlations in the center of systems as large as
$L=300$ along the legs and rungs in \Fig{fig:SU3_DMRG}(a) and
(b), respectively.  We can distinguish three regions.
The small $\Jp$ regime extending from $0^+$ to about $\Jp \sim
1 \equiv \Jca$ represents the $2\kf$ VBC.  The large $\Jp$
regime $\Jp>\jpc\equiv\Jcb$ describes the gapless $SU(3)_1$
WZWN phase. These two phases are separated by the intermediate
phase with $\Jp \in [\Jca, \Jcb ]$ [shaded areas in
\Fig{fig:SU3_DMRG}(a-b)] which corresponds to a gapless
incommensurate phase, as qualitatively suggested by the
preceding discussion.

The incommensurate behavior of the system is analyzed
in \Fig{fig:SU3_DMRG}(c) via spatial Fourier transform in
the leg direction of the rung bond strengths $\SdotSrung$.
The incommensurate wavevector $q(\Jp)$ is determined by local
maxima as traced by the red dots, which are further analyzed
in \Fig{fig:SU3_incommL}.

The upper boundary $\Jcb$ is extremely steady and
easily visible already within exact diagonalization of very
small system sizes, resulting in $\Jcb \approx 1.56$, [see
\App{fig:SU3_ED_Crossings}].  From DMRG simulations for
$L=60$ ladders, the value is $\Jcb=\jpc$ to within an
uncertainty of $+0.002$ towards a slightly larger value.
The sharpness of the upper boundary $\Jcb$ is due to vertical
slope in the incommensurate wavelength in \Fig{fig:SU3_DMRG}(c),
$ \bigl| \tfrac{d \qinc }{ d \Jp } \bigr\vert _{(\Jcb)^{-}}
\to \infty$.  In contrast, the lower boundary $\Jca$ is
strongly sensitive on system size, as it depends on whether or
not an incommensurate wave length still fits into a given
finite system.
Here for $L=60$, the transition to commensuration actually
already occurs around $\Jp\sim 1.2$ [\Fig{fig:SU3_DMRG}(c)].
For significantly larger systems this moves as low as
$\Jp\simeq 1.08$ (data not shown). In this sense,
$\Jca\approx1$ is an estimate from an extrapolation as shown
in the inset of \Fig{fig:SU3_DMRG}(b).
Now given that the slope $\tfrac{d \qinc}{d\Jp}$ tends to
zero for $\Jp \to (\Jca)^{+}$, one may question whether a
finite $\Jca>0$ exists at all.  However, this transition
at a finite $\Jca$ is confirmed  in \Fig{fig:SU3_gap}.
There we show that the \SU{3} ladder is gapped e.g. at
$\Jp=0.75$ in the thermodynamic limit and that, indeed,
this gap closes around $\Jp\sim1$.

The incommensurate behavior of the bond energies is tightly
linked to the global ground state parity $P$ as $L$ is varied.
As demonstrated in \Fig{fig:SU3_DMRG}(d), for every full
incommensurate period $\lambda/2$ [cf.~\Fig{fig:SU3_incommX}]
in the bond energies along a single long ladder, also the
parity of the ground state of a short ladder flips when
changing its length $L \to L+\lambda/2$ (black data).
This confirms the intuitive notion that by hosting a
symmetric rung state, this replaces an antisymmetric rung
state, and hence changes the parity.
As an aside, \Fig{fig:SU3_DMRG}(d) also shows that the
incommensurate behavior on bond energies along rungs completely
coincides with an incommensurate behavior also seen for the
bond energies along the legs.

Next we revisit the ED analysis of the $\nSxx$ profiles in
\Fig{fig:SU3_Panel-ED}, but now for significantly larger
system at $L=60$ based on DMRG data. With $\nSx = \SdotSrung +
\tfrac{2}{3}$ [cf.~\Eq{eq:parity:2}], this further analyzes
the data underlying \Fig{fig:SU3_DMRG}(b-d). 
The DMRG data is presented in \Fig{fig:SU3_incommD} for
various values of $\Jp$.  In each panel (a-g) we draw the bare
data (blue line), but also draw additional lines by connecting
every third data point to better visualize the incommensurate
behavior, using different lighter colors for different values
of (${x}{\mod 3}$).  Panel (a) is just below $\Jca(L=60)$,
having $\Jp=1.2$, whereas panel (g) is just above $\Jcb$.

\FIG{fig:SU3_incommD}(h) summarizes the actual expectation
values $\NSx$ vs. $\Jp$, where the data plotted,
$\nuS\equiv 3\NSx/L$, corresponds to the density of
symmetric rung states per $3$ rungs.  Note that at the
boundaries of the incommensurate phase (shaded area)
$\nuS$ is not strictly integer.  For example, just above
$\Jcb$, $\nuS \simeq 0.1$ is finite due to presence of
virtual excitations into the symmetric multiplet at higher
energies.  Just below $\Jca$, $\nuS \simeq 1.27$ is
somewhat above $1$ for the same reason.  In this sense,
as already pointed out in the ED discussion earlier,
the equivalence of $\NSx$ and $\Nm$ is not strict because
$\NSx$ also contains an inseparable contribution from
virtual excitations. Note that $\nuS$ is also not
constant within the VBC phase, since for $\Jp<\Jca$,
$\nuS$ necessarily increases still up to $\nuS=2$ at $\Jp=0$
which, matter of fact, is the expected value for two
decoupled chains.  Within the incommensurate region,
however, $\NSx$ closely follows the dependence of $m$.
The switching of the ground state parity vs. $\Jp$
for fixed system length $L$, on the other hand,
is an exactly countable feature within the numerical
simulations [e.g. see also \Fig{fig:SU3_incommX}(b)],
which thus makes $m$ a well-defined quantized integer.

\FIG{fig:SU3_incommD}(a) shows the CDW/VBC with a regular
period 3 pattern.  It has $\Nm=L/3=20$ ($\NSx=23.6$). Here,
due to finite size, the system is already locked into the
VBC phase. 
At the other end of the incommensurate regime,
panel (f) already belongs to the large $\Jp$ regime with
$\Nm=0$  ($\NSx=2.1$).  The small oscillations visible in
panel (g) are an order of magnitude smaller than in the VBC
phase [panel (a)], and in contrast to VBC phase, they decay
with system size.  When the ground state switches from the
$\nSxx$ profile in \Fig{fig:SU3_incommD}(g) down to the
panels (f) and (e), 
with each switch the expectation value $\NSx$ increases
by approximately 1, having $\Nm=0,1,2$, 
respectively, as indicated by $\Nm$ blue arrows
[the case $\Nm=1$ in panel (f) is very close to the phase
boundary $\Jcb$ hence its value of $\NSx$ is strongly
affected by the finite system size].
\FIG{fig:SU3_incommD}(d) already reaches $m=7$.

\FIG{fig:SU3_incommD}(c) down to (a) correspond to $\Nm=18,19$,
and $20$, respectively.  Here it is more challenging to
identify the location of these $m$
symmetric rung states. However one can spot the appearance
of $3\times (L/3-\Nm)$ domain walls between different local
patterns of the three degenerate CDW/VBC patterns~\cite{Su1981}
whose positions are indicated by the red arrows.
These 3 or 6 domain walls in \Fig{fig:SU3_incommD}(b) or
(c), respectively, are fractionalized defects introduced
into the CDW/VBC by reducing $\NSx$ by approximately 1 for
each switch in the $\nSxx$ profile when going from
\FIG{fig:SU3_incommD}(a) to (b), or from (b) to (c).
It thus seems that one symmetric rung state defect doped
into the CDW/VBC decays into three  domain walls, a hallmark
of fractionalization~\cite{Su1981}.

In summary, the incommensurate wave length tracked by the
red dots in \Fig{fig:SU3_DMRG}(c) can be rationalized as
$q(\Jp)=\pm \tfrac{2\pi}{L} \Nm(\Jp)$, where the integer
$\Nm$ \mbox{$(\sim{\NSx})$} denotes e.g.  the number of level
crossings discussed above.  \FIG{fig:SU3_DMRG}(c) highlights
the Lifshitz-like transition occurring at $\Jcb$, when the
first symmetric rung states start to populate the ladder
upon lowering $\Jp$. In the intermediate phase, $\Jp$
controls the incommensuration in a way reminiscent of
magnetization curves in the vicinity of magnetization
plateaux or the density controlled by a chemical potential
in the vicinity of a quadratic band edge.
The difficulty in precisely locating the lower boundary of
the incommensurate gapless phase seems to be tied to the
nature of the period-3 CDW/VBC which neighbors the
incommensurate phase.  It has been observed in a number of
one-dimensional CDW-like systems, that the doping of the
CDW can lead to a fractionalization of the defects,
leading to a seemingly slow onset of the incommensuration
with the field or chemical potential, see
e.g.~Refs.~\cite{Yang1966,Okunishi2003,Fouet2006}.

\subsubsection{Intermediate Phase for $N=5$}


\begin{figure}[tb]
\includegraphics[width=\linewidth,clip]{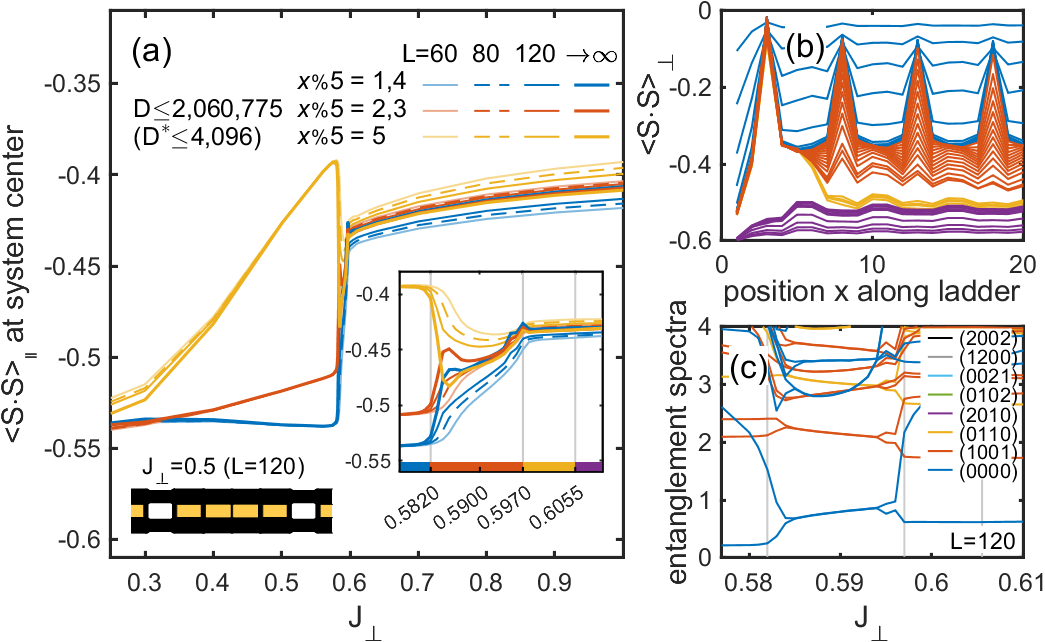}
\caption{(Color online) \SU{5} Heisenberg ladder ---
  Panel (a) Similar analysis as in \Fig{fig:SU46}(a),
  yet grouping data w.r.t. $x \mod 5$ (see
  legend).  This results in three branches for $\Jp <
  \jpa\equiv \Jca$ due to pentamerization, as shown in the
  lower left inset for $\Jp=0.5$.  From our numerical data
  we conclude that this gapped phase persists down to
  $\Jp\to0^+$.  The lower right inset shows a zoom into
  the extremely narrow intermediate phase for
  $\Jp\in[\jpa,\jpb]$.
  Panel (b) Nearest-neighbor $\SdotSrung^{\mathrm{NN}}$
  rung data at the open left boundary of the system. By
  color coding the ranges in $\Jp$ as with the x-axis in
  the lower right inset of panel (a), the intermediate
  phase is linked to the sharp emergence of an edge
  feature at the open boundary of the ladder.
  Panel (c) Analysis of the entanglement flow for the same
  window in $\Jp$ as the right inset in (a). These
  entanglement spectra were computed  for fixed $\Jp$
  in the center of uniform length $L=120$ ladders
  Lines with the same color belong
  to the same \SU{5} symmetry sector, using a compact
  Dynkin labeling scheme in the legend (see \App{app:dmrg}).
  The results are also
  consistent with DMRG scans in $\Jp$ (e.g. see Fig.~\ref{fig:SU5_scan} 
  in \App{app:N5}).
}
\label{fig:SU5_DMRG}
\end{figure}


The same symmetry considerations regarding the
rung parity change coming from the large $\Jp$ regime to
the small $\Jp$ VBC regime apply to $N=5$. Whereas in the $N=3$
case this change occurs gradually in a rather extended
$\Jp$ region, the case $N=5$ is quite different. At first
sight the transition seems to happen quite abruptly around
$\Jp\sim0.58$, however a careful DMRG investigation shows
that the critical, large $\Jp$ and the small $\Jp$ VBC
phases are separated by a finite (albeit rather tiny)
intermediate phase, thus avoiding a direct first order
transition. We present a compact overview of these
findings in this subsection and provide in-depth material
in \App{app:N5}. 

Our results for the \SU{5} ladder are summarized in
\Fig{fig:SU5_DMRG}.  It features an extremely narrow
intermediate phase that stretches over a less than 3\%
variation of $\Jp \in [\jpa, \jpb]$. Yet it is stable,
in the sense that its window slightly increases with
increasing system size [inset to \Fig{fig:SU5_DMRG}(a)].
Importantly, it smoothly connects strongly different
values of $\SdotSleg$ for $\Jp<\Jca$ as compared to
$\Jp>\Jcb$. Consequently, what would have been a first
order transition in the absence of the intermediate
phase, become two (possibly) second-order phase
transitions.

In stark contrast to the \SU{3} case, however, the
intermediate phase of the \SU{5} ladder shows no signs
of incommensuration in the nearest-neighbor
spin-spin correlations [e.g. see smooth behavior of the
nearest-neighbor $\SdotS$ data across entire intermediate
phase in \Fig{fig:SU5_scan}(c)]. Instead, the data for the
intermediate phase in \Fig{fig:SU5_DMRG} show properties
that are typically rather associated with topological
phases:
(i) The multiplets in the entanglement spectra pair up
extremely systematically, as shown in \Fig{fig:SU5_DMRG}(c)
[see also \Fig{fig:SU5_scan}(d)].
(ii) A striking edge feature snaps into the system when
approaching the intermediate phase from large $\Jp$.
(iii) Finite size extrapolation in the bond strengths
reduces the strength in pentamerization, which may points
towards a uniform isotropic ladder in the thermodynamic
limit [see inset to \Fig{fig:SU5_DMRG}(a)].

The edge feature, as shown in \Fig{fig:SU5_DMRG}(b), 
already occurs for a value of $\Jp$ slightly outside
$\Jcb$, and therefore initially it appears as a pure boundary
feature. This fact alone could be interpreted as a
precursor of a first-order transition due to open boundary
conditions~\cite{Fouet2006b}.  Nevertheless, once $\Jp$
crosses $\Jcb$, the edge feature still persists within the
intermediate phase where the bulk properties of the system
change profoundly.  More supporting data is also provided
in \App{app:N5}.

Furthermore, while the nearest-neighbor spin-spin
correlations in the case of the \SU{5} ladders are
commensurate, there are traces of incommensuration, e.g.
seen in the minor wiggles in the entanglement spectra in
\Fig{fig:SU5_DMRG}(c) just below $\Jcb$ [with more
resolution shown in \Fig{fig:SU5_scan}(d) and its inset].
This incommensuration also appears reflected e.g. in the
non-local expectation values of the block-parity
[\Fig{fig:SU5_comm}(a)].  Nevertheless, in either case,
the incommensuration decays quickly when going into the
bulk of the ladder [e.g. see \Fig{fig:SU5_comm}(b)].

A simple intuitive picture, however, that accounts for all
the properties seen in this newly uncovered intermediate
phase in the \SU{5} ladders is still lacking.  It shall
also be noted that for Heisenberg ladders with a
well-defined local multiplet, a generalized Lieb-Schultz-%
Mattis (LSM) theorem applies \cite{LSM61,Affleck1986,
Affleck1988}.  As a consequence, our two-leg ladder
in the defining representation of \SU{N} either
spontaneously breaks translation symmetry, or necessarily
possesses soft modes at momentum $4\pi/N$~%
\footnote{private communications with Ying Ran
(Boston) and Keisuke Totsuka (Kyoto).}
and hence a gapless spectrum.  This is found, for
instance, in the small and large $J_\perp$ regions for
$N=3$ and $N=5$.  In the intermediate phase for $N=5$,
however, we clearly see a two-fold degenerate ground state
and systematic two-fold degeneracy in the entanglement
spectra, all of which is tightly linked to parity and thus
apparently decoupled from the \SU{5} symmetry.
The two-fold degeneracy may be related to the presence of
the open boundaries. A more detailed study based on
periodic boundary conditions, however, is currently not
feasible numerically, given that within the intermediate
phase the \SU{5} ladders up to length $L=30$ with periodic
boundary conditions are still too strongly affected by
finite size effects. For the same reason, even when using
open boundary conditions, we cannot make definitive
statements on whether or not the ground state in the
\SU{5} intermediate phase is translationally uniform
in the thermodynamic limit.

\section{Conclusion and Outlook} \label{sec:outlook}

We analyze the \SU{N} Heisenberg ladder for
antiferromagnetic $\Jp$ with relevance for materials
\cite{Dagotto96, Caron11, Luo13, Schmidinger13} as well as
experiments in ultracold atoms \cite{Jordens08, Lake10,
Ward13-qsim, Cazalilla14, Zhang14, Scazza14, Cappellini14,
Pagano14, Bloch15}.  We uncover rich phase diagrams that
strongly depend on the parity of $N$. For even $N \leq 6$,
the spectrum is always gapped and the ground state is a
$4\kf$ VBC.  For odd $N$, we find a $2\kf$ gapped VBC
phase at weak coupling, a critical phase at strong
coupling, and an intermediate phase that is incommensurate
for $N=3$, but extremely narrow and reminiscent of
topological phases for $N=5$.  While topological
phases of the type encountered can be ruled out for a
single chain~\cite{Duivenvoorden2013,Capponi16}, it is
emphasized that the present system of a 2-leg ladder has
an additional discrete parity symmetry that is playing
an essential role.

We focused on the Mott insulating regime of symmetric
multi-flavor fermionic Hubbard models at filling of one
particle per site. Our detailed analysis vs. the number of
flavors $N$ already uncovers a remarkable phase diagram,
which reflects the complexity of strongly-correlated
quantum many-body phenomena, in general. The full
exploration of the original fermionic Hubbard model for
$N>2$ flavors, however, e.g. at different integer fillings
as well as doped systems for arbitrary chemical potential
and the full range of interaction strength $U$ remains a
wide, largely unexplored, yet exciting open field for
future research.
From a different perspective, our model with strong
ferromagnetic rung will map onto an effective chain with
symmetric ${\tiny \yng(2)}$ irrep of \SU{N}, for which
there exists a generalized Haldane
conjecture~\cite{Rachel09} and recent progress was made to
understand such symmetric irreps for $N=3$~\cite{Lajko17}.
More generally, larger representations are a simple way to
stabilize more exotic phases such as chiral
ones~\cite{Hermele11}. Another direct extension of this
work would be to study similar geometries such as zigzag
ladders, which can be realized
experimentally~\cite{Anisimovas16} where frustration is
expected to play an important role.  By focusing on
symmetric models, however, the full exploitation of the
underlying non-abelian symmetries offers the prospect of a
significant advance in terms of exact numerical
simulations on these systems.

\acknowledgments

The authors are grateful to V. Gurarie, M. Hermele, A.
Honecker, S.  Manmana, G. Mussardo, K. Totsuka and A.
Wietek for useful discussions.  The authors would like to
thank Yukawa Institute (Kyoto, Japan) for hospitality
during work on this project. 
AMT was funded by US DOE under contract number
DE-SC0012704. 
AW was also funded by the German
Research foundation, DFG WE4819/2-1 and DFG WE4819/3-1
until December 2017, and by DE-SC0012704 since.
AML was supported by the Austrian Science Foundation (FWF)
through projects I-2868 and F-4018.
PL and SC would like to thank CNRS (France) for financial
support (PICS grant).
This work was performed using HPC resources from CALMIP and
GENCI-IDRIS (Grant No. x2016050225 and No. A0010500225),
as well as the Arnold Sommerfeld Center, Munich.


\bibliographystyle{apsrev4-1}
\bibliography{sunbib} 
 
\appendix
\renewcommand\thefigure{\thesection.\arabic{figure}}
\setcounter{figure}{0}

\clearpage 

\section{DMRG simulations of SU(N) ladders} \label{app:dmrg}

The DMRG \cite{White92,Schollwoeck05,Schollwoeck11}
data in the main text as well as in these appendices 
fully exploit the underlying \SU{N} symmetry based on the
QSpace tensor library \cite{Wb12_SUN}.  While QSpace can
deal with general non-abelian symmetries, the discussion
in the following is constrained to \SU{N}.  By operating
on the level of irreducible representations (ireps) rather
than individual states, by construction, symmetry
multiplets remain intact.  Conversely, from a practical
point of view, by switching from a state-spaced
description to a multiplet based description, e.g. for the
case of \SU{5}, this allows us to effectively keep beyond a
million of states, and therefore to go significantly
beyond the current state-of-the-art of DMRG simulations.


A ladder of length $L$ is represented by an $2\times L$
lattice, and hence consists of $L$ rungs and $2L$ sites.
The DMRG simulation proceeds along a snake-like pattern
from rung to rung, and therefore operators on a chain of
$2L$ sites.  While this makes the algorithm more
efficient, this does not explicitly include the parity
symmetry w.r.t. exchange of upper and lower leg.
Throughout we use the term {\it parity symmetry} to refer
to this reflection symmetry, nevertheless.  All energies
are taken in units of $J_{||} := 1$, with $J_\perp \geq 0$.

DMRG simulations can also be used to scan parameter
regimes by (slowly) tuning parameters in the Hamiltonian
(here $\Jp$) e.g. as a linear function of the ladder
position $x$ within a single DMRG run\cite{Zhenyue15}.
This is referred to as a \emph{DMRG scan}. In the
resulting plots we directly show the underlying tuned
parameter $\Jp$ rather than the ladder position $x$. To be
explicit, we may also may write the label of
the horizontal figure axis e.g. as ``{\it position $x$ along
the ladder $\to \Jp$}''.  These scans serve the purpose of
quick snapshots of the physical low-energy behavior along
a line in parameter space with blurred open boundary
condition.  Eventually, these calculations are typically
followed by DMRG simulations of uniform systems at specific
parameter points of interest.

Symmetry labels for \SU{N} multiplets are specified by
their Dynkin labels $q\equiv (q_1,\ldots,q_{\rsym})$,
with $\rsym=N-1$ the rank of the symmetry \SU{N} and $q_i
\in \mathbb{N}_0$ non-negative integers \cite{Elliott79,
Gilmore06, Wb12_SUN}.
For example, the Dynkin label for an \SU{2} multiplet of
spin $S$ is given by $q=(q_1)$, with $q_1=2S$ the number
of boxes in its Young tableau. For general \SU{N},
the Dynkin labels count the differences from one row of a
Young tableaux to the next, with the total number of boxes
in the corresponding Young tableau $Y_q$ given by $n_Y \equiv
\sum_{k=1}^{N-1} k\cdot q_i$.

Throughout, a compact label format is adopted that simply
skips commata and spaces in a set of Dynkin labels.
For example, $(10) \equiv (1,0)$ is the defining
representation of \SU{3} of dimension $d=3$.  If the
labels $q_i$ have values beyond $9$, the counting is
continued with alphabetic characters $A-Z$ [e.g. $(1A)
\equiv (1,10)$ akin to hexadecimal digits].  This suffices
for all practical purposes since no labels with values
$q_i > Z\equiv 35$ are encountered.  

When convenient and unique, the alternative convention of
specifying a multiplet simply by its dimension ${\bf d}$
is adopted.  For example, for \SU{3} the defining
representation is given by $\mathbf{3} \equiv (10)$. This
customary labeling scheme becomes ambiguous, though, and
thus fails if different non-dual multiplets share the same
multiplet dimension [e.g. for \SU{3}, the ireps (40) and
(21) have the same dimension $d=15$].
The labeling scheme by dimension (as far as applicable)
uses bold font to emphasize that a multiplet typically
represents multiple symmetric states. These multiplets can
quickly become ``fat'' for larger \SU{N}. In practice,
they quickly reach dimensions up to $d \lesssim
10^{\rsym}$, again with $\rsym=N-1$ the rank of the
symmetry \SU{N}.  That is while \SU{2} typical encounters
multiplets that only have up to $\lesssim 10$ states (i.e.
spin $S\lesssim5$), for \SU{5}, in fact, this quickly
reaches up to and beyond $d=10,000$ states for a single
multiplet!
 
The dual or ``barred''-representation $\bar{q}$ always
shares exactly the same dimension $d=|q|$ 
in terms of number of states as the original irep $q$.  In
Dynkin labels, the dual irep in \SU{N} is simply given by
reversed labels, i.e.  $\bar{q} = (q_{N-1},\ldots,q_{1})$.
For example, the dual to the defining irep is given by
$\bar{\bf N} \equiv (0\ldots01)$. While
$|\bar{{\bf d}}| = |{\bf d}|$, by
convention, $\bar{{\bf d}} \le {\bf d}$ in lexicographic
order on their corresponding Dynkin labels, where equality
holds for self-dual ireps.
 
The spin operator always transforms in the adjoint
representation of dimension $d=N^2-1$ which simply
represents the set of all generators of the Lie algebra.
It is obtained whenever a non-scalar multiplet $q$ is
combined with its dual $\bar{q}$.  For example, taking the
defining representation, ${\bf N} \otimes \bar{\bf N}
\equiv (10\ldots0) \otimes (0\ldots01) = 1 +
(10\ldots01)$, with $1 \equiv (0\ldots0)$ the scalar
representation, one arrives at the symmetry labels of the
spin operator for \SU{N>2}, given by $q_S=(10\ldots01)$.
For \SU{2}, one has $(1) \otimes (1) = (0) + (2)$ and
therefore $q_S=(2)$, i.e. $S=1$.

Even though we consider general \SU{N} symmetries,
we will nevertheless use the familiar \SU{2} language
where convenient and possible.  That is, the \SU{2} spin
operator $\hat{\mathbf{S}}$ is generalized to the \SU{N}
``spin'' operator where $\hat{S}^A$ are the \SU{N} spinor
components with $A=1,\ldots,N^2-1$, e.g. as used in
\Eq{eq:H:spinladder} or \Eq{eq:spinop} in the main text.
In particular, we also refer to $S\cdot S$ as the
(generalized) spin-spin interactions.
For example, $\SdotSleg$ will refer to the
nearest-neighbor $S \cdot S$  expectation values along the
legs in the DMRG ground state $| 0 \rangle$, i.e.
%
%
\begin{eqnarray}
   \SdotSleg &\equiv& \langle 0|
   \hat{\mathbf{S}}_{\ell x} \cdot
   \hat{\mathbf{S}}_{\ell,x+1} |0\rangle \equiv \sum_A
   \langle 0 | \hat{S}^A_{\ell x}\, \hat{S}^A_{\ell,x+1} |0\rangle
\text{,}\ \ \qquad \label{eq:SSleg}
\end{eqnarray}
For the \SU{N} ladder systems considered here, these
expectation values along the legs are always the same for
either of the two legs $\ell=1,2$ (``upper'' and ``lower''
leg), hence the leg-index $\ell$ can be dropped [i.e.
without implicit summation over $\ell$ in \Eq{eq:SSleg}].
The index $x$, finally, refers to the position along the
ladder with lattice spacing $a=1$. Hence $x\pm1$ refers to
nearest neighbor rungs. Depending on the context, $x$ may
also refer to the bond in between rungs $x$ and $x+1$.

Similarly, $\SdotSrung$ refers to nearest-neighbor spin
correlators across a single rung, i.e.
\begin{eqnarray}
   \SdotSrung &\equiv& \langle 0|
   \hat{\mathbf{S}}_{1 x} \cdot \hat{\mathbf{S}}_{2 x} |0\rangle
   \equiv
   \sum_A \langle 0 | \hat{S}^A_{1 x}\, \hat{S}^A_{2 x} |0\rangle
\text{.}\ \ \qquad \label{eq:SSrung}
\end{eqnarray}
%
%
In general, $\SdotSleg$ and $\SdotSrung$ depend on
the position $x$.

The Heisenberg ladders under consideration have the
defining irep at each site. For this specific case, the
scalar operators $\mathbf{S}_i \cdot \mathbf{S}_j$ acting
e.g. on nearest-neighbor sites $i \equiv (\ell x)$ and $j
\equiv (\ell' x')$ can be simply related to the parity
$\hat{P}_{i,j}$ which swaps the states of the two sites.
In particular then, it holds for general \SU{N} for two
sites $i \neq j$,
\begin{equation}
  \hat{P}_{ij} = \tfrac{1}{N} + 2\, \mathbf{S}_i \cdot \mathbf{S}_j
\text{ .}\label{eq:P2SS}
\end{equation}
This can be derived from elementary symmetry considerations
as follows. Given that we only consider the defining
representation $q$ of \SU{N} on each site, from the
product $q \otimes \bar{q}$ we see that aside from the
trivial identity operator $\hat{1}$, the only non-trivial
operators that can act on the state space of a single site
$i$ are the components of the spin operator $\hat{\mathbf{S}}_i$.
Consequently, the only non-trivial scalar operator that
can act across two sites $i$ and $j$ as part of an
interaction term in a Hamiltonian is the spin-spin
interaction $\mathbf{S}_i \cdot \mathbf{S}_j$.
On the other hand, the permutation $\hat{P}_{ij}$ (i.e.
swap) of two sites $i$ and $j$ is also a non-trivial
hermitian operator that can equally well serve as an
individual term in an \SU{N} symmetric Hamiltonian.  From
the above elementary symmetry considerations, therefore it
follows that $\hat{P}_{ij}$ must be a combination of the
trivial operator and spin interaction, i.e.  $\hat{P}_{ij}
= a\,\hat{1}_{i j} + b\, \mathbf{S}_i \cdot \mathbf{S}_j$,
with coefficients $a$ and $b$ to be determined.
Clearly, with the spin operators being traceless,
then with the normalization convention
$\trace{S_i^A S_i^B}=\tfrac{1}{2} \delta_{AB} $
[cf.~\Eq{eq:spinop}],
it must hold within the state space of two defining
representations of \SU{N} at sites $i$ and $j$ that
$N = \trace{\hat{P}_{i j}} = a N^2$, and hence
$a=\tfrac{1}{N}$.  Furthermore, with $\hat{P}_{i j}^2 =
\hat{1}_{i j}$, it follows $N^2 = \trace{\hat{P}_{i j}^2}
= a^2\, \trace{\hat{1}_{i j}} + b^2\, \trace{
(\mathbf{S}_i \cdot \mathbf{S}_j )^2 } = 1 + b^2 \,
(N^2-1)\bigl(\tfrac{1}{2}\bigr)^2$, and therefore $b=2$
which proves \Eq{eq:P2SS}.

\section{Further exact diagonalization results \\
  on the $N=3$ intermediate phase}
\label{app:ed_details}

As a first attempt to check the phenomenological picture
for the $N=3$ intermediate phase, namely that starting
from strong coupling where the rungs are predominantly
antisymmetric, we want to understand the effect of
decreasing $J_\perp$ in terms of symmetric rung density.
We have performed Exact Diagonalization (ED) using
Lanczos algorithm on ladders $2\times L$ ($L$ is chosen
as a multiple of 3 to accomodate a singlet state as well
as the VBC expected at weak coupling) with open boundary
conditions (OBC) using global color conservation. We can
also use the global parity, that exchanges the two legs,
as a discrete symmetry, and hence label the ground-state
as even or odd. 

\begin{figure}[tbp]
\includegraphics[width=0.95\linewidth]{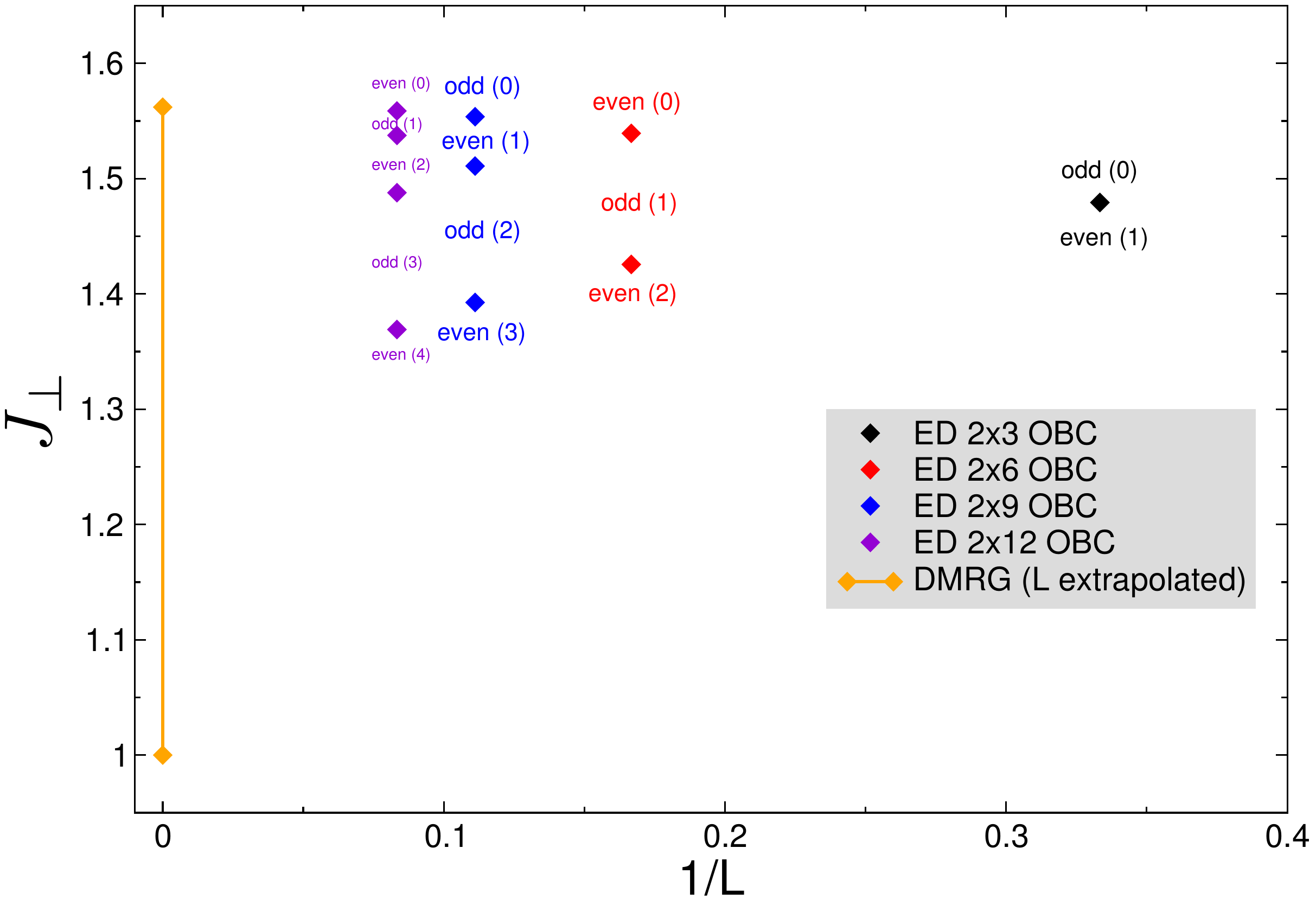}
\caption{(Color online)
  Locations of the changes in the ground-state parity
  obtained from ED results on \SU{3} Heisenberg $2\times
  L$ ladder as  a function of $1/L$.  The number $(\NS)$
  indicates the number of ground state level crossings
  encountered coming from large $\Jp$ (see
  \Fig{fig:SU3_Panel-ED} in the main text). There are
  $L/3$ level crossings distributed over a significant
  region of $J_\perp$, compatible with the extension of
  the intermediate phase found in DMRG.
} 
\label{fig:SU3_ED_Crossings}
\end{figure}

From the strong coupling picture at large $J_\perp$
where all rungs are antisymmetric, it is clear that the
ground-state should have parity $(-1)^L$. In
Fig.~\ref{fig:SU3_ED_Crossings}, we do confirm this
statement, but we also notice that, while decreasing
$J_\perp$, there is a sequence of level crossings: on a
$2\times (3p)$ ladder, there are $p$ changes in the
ground-state parity. Moreover, the range of parameters for
these transitions is quite wide, in good agreement with
the extension of the incommensurate phase found in DMRG
simulations.

Overall, these ED data suggest that there is no sharp
first order transition, but rather a finite intermediate
region.


\section{Further DMRG results
\protect\newline\mbox{\ }\hspace{-0.5in}
on the $N=3$ intermediate phases} \label{app:N3}

\begin{figure}[tbp]
\begin{center}
\includegraphics[width=0.95\linewidth]{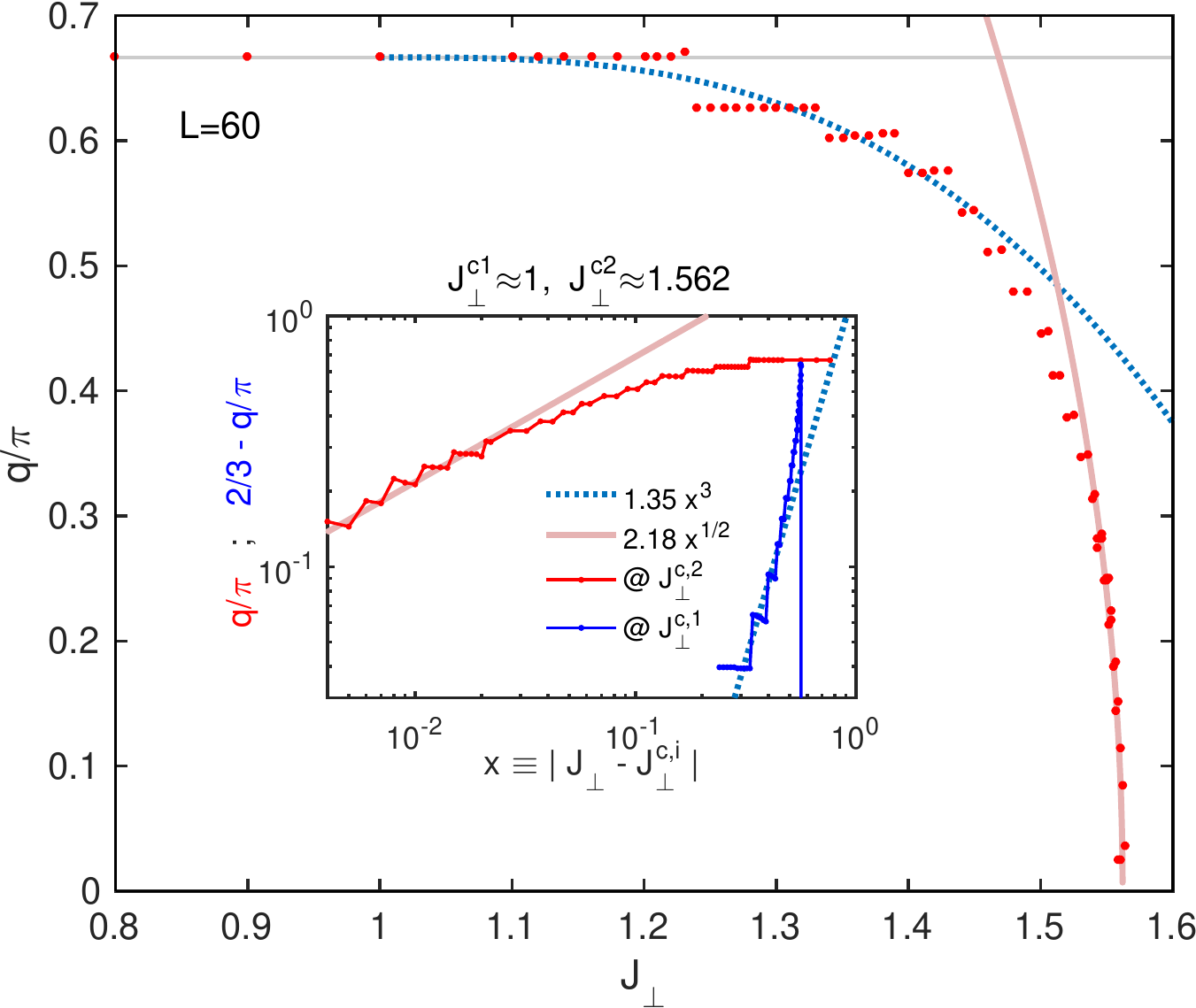}
\end{center}
\caption{(Color online) 
   Incommensurate wave length in uniform SU(3) Heisenberg
   ladders of length $L=60$ with open boundaries in the
   incommensurate regime $\Jp \in [\Jca,\Jcb] \equiv
   [{\sim}~1, \jpc ]$ [data taken from
   \Fig{fig:SU3_DMRG}(c) in the main paper, where red
   points trace the incommensurate maxima in Fourier
   space].  The inset suggests a scaling for the
   incommensurate wave vector that is
   $\bigl(\tfrac{2\pi}{3} - \qinc\bigr) \propto
   |\Jp-\Jca|^{\sim 3}$ close to $\Jca$, and $\qinc
   \propto \sqrt{|\Jp-\Jcb|}$ close to $\Jcb$.
}
\label{fig:SU3_incommL}
\end{figure}


\begin{figure*}[tbp]
\begin{center}
  \includegraphics[width=0.85\linewidth]{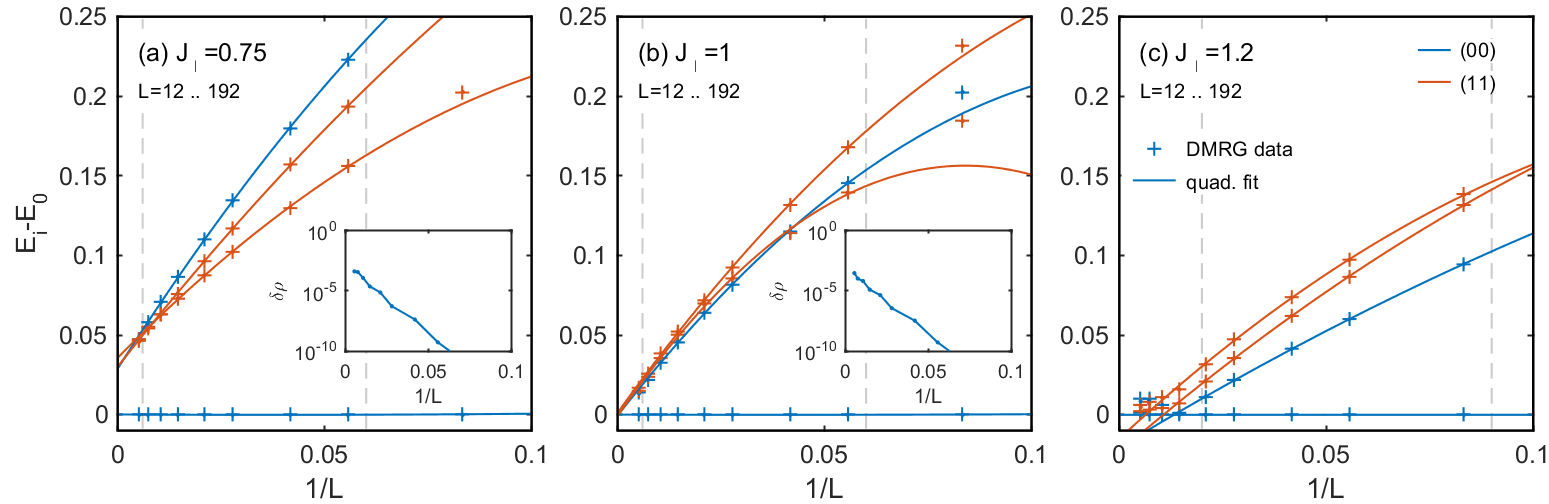}
\end{center}
\caption{
   Analysis of excited states across the lower boundary
   $\Jca$ of the incommensurate phase for the \SU{3}
   Heisenberg ladder. Panel (a) for $\Jp=0.75$ shows a
   small but finite energy gap when extrapolating to the
   thermodynamic limit $1/L\to0$.  Panel (b) at $\Jp=1$
   suggests the onset of criticality, whereas for panel
   (c) at $\Jp=1.2$ due to the presence of incommensurate
   correlations other states take over the ground state
   for sufficiently large systems (see crossing of the
   lowest blue line, which served as energy reference).
   For the simulations here, several lowest system
   eigenstates are targeted simultaneously within a single
   DMRG run for uniform systems up to length $L=96$ using
   open boundary conditions.  The colors encode symmetry
   sectors, which shows that two multiplets were included
   in the $q=(00)\equiv \mathbf{1}$ sector (singlets; blue
   data points) as well as two multiplets in the adjoint
   representation $q=(11) \equiv \mathbf{8}$ (orange
   data points).  Therefore while only including $4$ multiplets
   total, this DMRG simulations actually targeted the
   lowest 18 states.
   The insets of panels (a-b) show the maximum
   discarded weight within DMRG in the last sweep
   while simultaneously targeting the low-energy multiplets
   above. For all ladders at least $1024$ multiplets
   ($>51,000$ states) were kept. For $L\leq 69$,
   at least $2048$ multiplets ($> 111,000$ states) were kept.
   The gray vertical lines indicate the region
   that was used to fit the data (crosses)
   to the thermodynamic limit $L\to\infty$ using a plain
   quadratic polynomial fit (solid lines).
}
\label{fig:SU3_gap}
\end{figure*}


\begin{figure*}[tbp]
\begin{center}
  \includegraphics[width=0.80\linewidth]{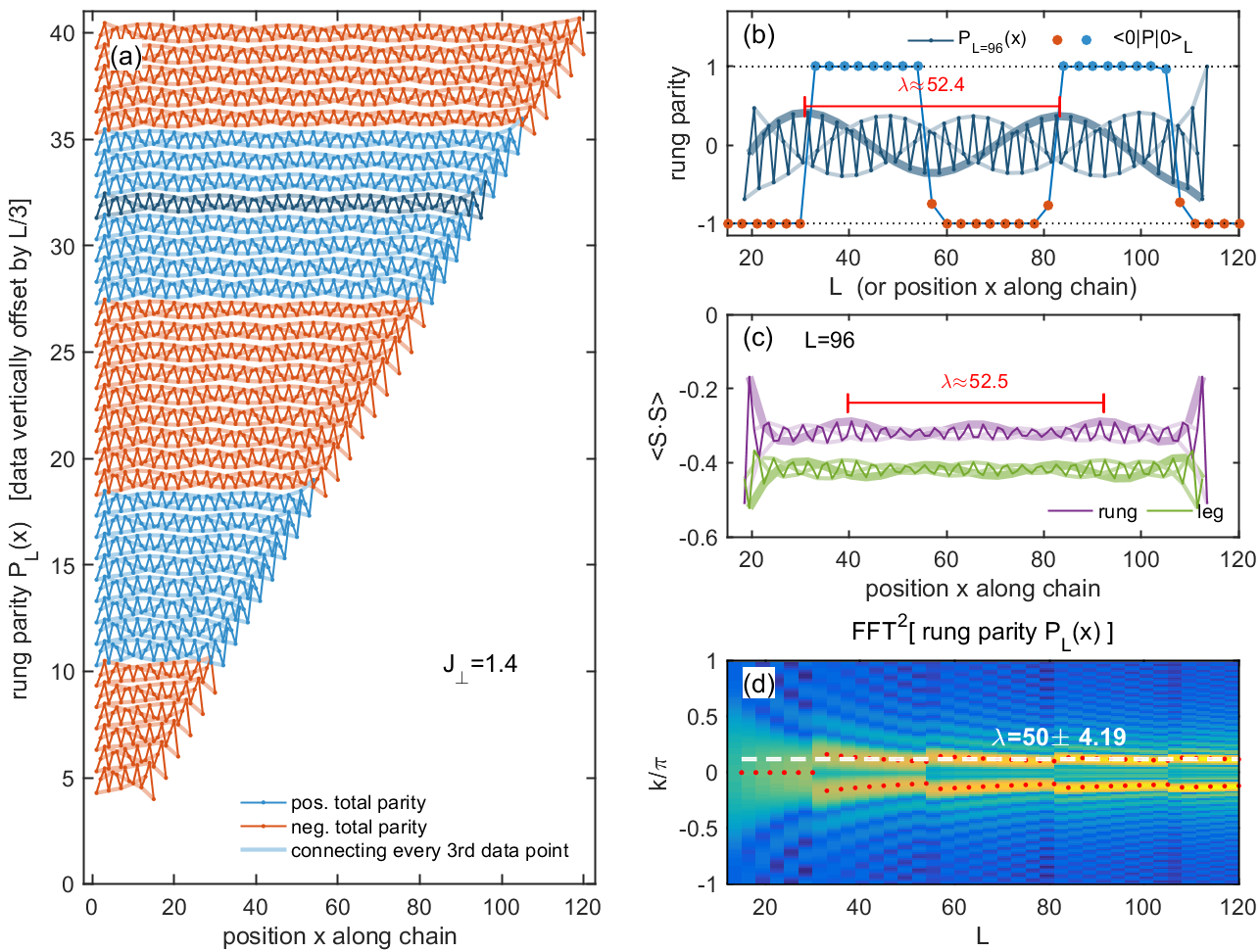}
\end{center}
\caption{
   Analysis of the incommensurate phase in the \SU{3}
   Heisenberg ladder for $\Jp=1.4$.
   Panel (a) shows the block parity $P_L(x) \in [-1,1]$ of
   the ground state [cf.~\Eq{eq:incommX}], where the data
   is vertically offset by the integers $L/3$ for different system
   sizes $L=15,18,\ldots,120$.  The incommensurate wave is
   highlighted by simply also connecting only every
   third data point (light colors).  The total rung
   parity, i.e. the last data point $P_L(L)$ of each curve
   determines whether the full ground state is symmetric
   (blue) or antisymmetric (orange) under rung exchange.
   Panel (b) compares the block parity $P_{L}(x)$ for a
   single system of fixed length $L=96$ [data horizontally
   offset by $dx=17.5$; data also marked in darker colors
   in (a)] to the full parity $P(L)$ for systems vs.
   varying length $L$ (dots) [this corresponds to the last
   data points $P_L(L)$ of each curve in (a)]. The
   resulting incommensurate period of $P_{L}(x)$ clearly
   matches the period seen in $P(L)$.
   For comparison, panel (c) shows the incommensurate
   behavior seen in the $\SdotS$ data [similar to panel
   (b), again for $L=96$, and data horizontally offset].
   The incommensurate wave length is the same up to a
   factor of $2$ (see text).
   Panel (d) shows the Fourier transform of the data in
   panel (a). In order to focus on the incommensurate
   behavior, this is the norm-squared of the FFT data when
   including rungs $x=i,i+3,i+6,\ldots$, subsequently
   averaged over $i=1,2,3$. The incommensurate wavelength
   $\lambda$ is obtained by tracking the peaks in the FFT
   data (red dots).  Finite-size effects lead to
   diminishing jumps in $\lambda$ as $L$ increases. For
   the standard deviation $\delta\lambda$ indicated in the
   panel, the data was taken in the range $L \in [80,120]$ 
}
\label{fig:SU3_incommX}
\end{figure*}


\FIG{fig:SU3_incommL} analyses the behavior of the
incommensurate wavelength $q(\Jp)$ close to the phase
boundaries of the incommensurate regime.  This suggests a
square-root like onset of $q$ close to $\Jcb$, and a power
law with an exponent $\alpha\gtrsim3$ at $\Jca$.

Since the slope $\tfrac{d \qinc}{d\Jp}$ tends to zero for
$\Jp \to (\Jca)^{+}$, one may question whether a finite
$\Jca>0$ exists at all.  However, as we demonstrate in
\Fig{fig:SU3_gap} by explicitly computing several excited
states, $\Jp=0.75$, indeed, possesses an energy gap in the
thermodynamic limit (panel a). Note that, although the VBC
state is 3-fold degenerate in the thermodynamic limit (or
with periodic boundary conditions), the ground-state is
unique here due to open boundary conditions that select a
particular VBC order.  This gap then closes around
$\Jp\simeq \Jca \approx  1.,$ (panel b), with a set of
crossing ground states within the incommensurate region as
the system size increases ($\Jp=1.2$; panel c).

Note also that for $\Jp<\Jca$ the first excited singlet
state (upper blue line) lies above the lowest excited
spinful state (orange lines) in the adjoint representation
$(11)$ [panel (a)]. For $\Jp>\Jca$, however this blue line
crosses below the orange lines, such that for system sizes
that are too short to harbor an incommensurate wavelength,
the first excited state is also a singlet [e.g. panel (c)
for $1/L \gtrsim 0.014$, which compares well to the crude
fit of the incommensurate wavelength $\lambda$ in
\Fig{fig:SU3_incommL} for $L=60$, having
$\tfrac{2}{\lambda} \equiv \tfrac{2}{3} - \qinc/\pi
\simeq 0.011$ for $\Jp=1.2$].

As explained in main text, \Sec{sec:SU3_incomm}, the
incommensurate regime is characterized by a rapid
succession of ground states with changing parity when
varying $\Jp$. Equivalently, for fixed $\Jp$ the ground
state parity changes with system length.  As shown in
\Fig{fig:SU3_incommX}, the resulting length scale is
directly related to the incommensurate wave length
observed in the $\SdotS$ data.
Specifically, \Fig{fig:SU3_incommX}(a) shows data
for the block parity of the ground state,
\begin{eqnarray}
   P_L(x) \equiv \langle 0| \prod_{x'=1}^{x} \hat{P}_{x'} |0\rangle
\text{ .}\label{eq:incommX}
\end{eqnarray}
With $\hat{P}_{x}$ the swap operator of the two sites at
rung $x$ [cf.~\Eq{eq:P2SS}], this swaps the legs of the
ladder for all rungs $x'\leq x$ for fixed $L$.  Therefore
$P_L(L) \equiv \langle 0| \hat{P} |0\rangle_L$ is the full
rung parity of a given DMRG ground state which also
determines the color of the data in
\Fig{fig:SU3_incommX}(a). The switching of the global
parity vs. system length $L$ is consistent with the
incommensurate pattern seen in the block parity $P_L(x)$
[\Fig{fig:SU3_incommX}(b)].  Up to a factor of $2$, this
wavelength $\lambda$ is also consistent with the one
observed in the $\SdotS$ data [\Fig{fig:SU3_incommX}(c)].
The additional factor of $2$ here suggests that the $\SdotS$
data simply cannot differentiate the sign of the block
parity $P_L(x)$.

\FIG{fig:SU3_incommX}(d) plots the data in panel (a) in
Fourier space with focus on the incommensurate wave by
taking period-3 data subsets.  The fixed system size
results in slight squeezing of the incommensurate wave.
Finite-size jumps occur whenever another half-wavelength
fits in the system [see \Fig{fig:SU3_incommX}(a)].
They diminish in the limit $L \gg \lambda$.

\section{Further DMRG results
\protect\newline\mbox{\ }\hspace{-0.5in}
on the $N=5$ intermediate phase}
\label{app:N5}

\begin{figure*}[tbp]
\begin{center}
  \includegraphics[width=0.85\linewidth]{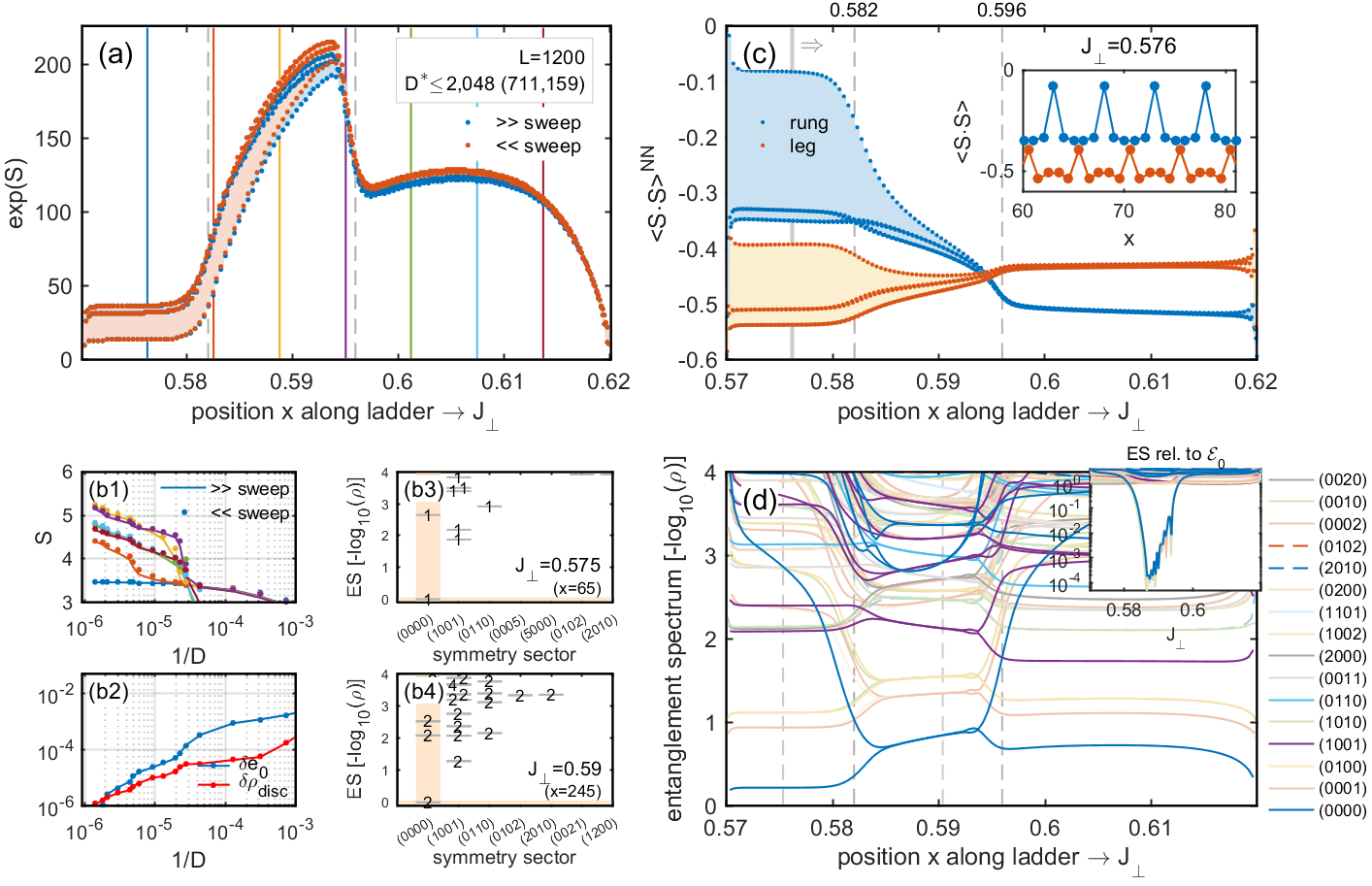}
\end{center}
\caption{
  DMRG scan across the narrow intermediate phase of the
  \SU{5} ladder [$\Jp \in [0.57,0.62]$ tuned linearly
  along a length $L=1200$ ladder with open boundaries,
  keeping up to $D^{\ast}\leq 2048$ multiplets ($D \leq
  711,159$ states)] ---
  Panel (a) shows the block entanglement entropy $S$ along
  the system. The intermediate phase $\Jp \in [\Jca,\Jcb]
  = [\jpa, \jpb ]$ is marked by vertical dashed lines
  [similar in panel (c-d)].  The data points are also
  connected by a line of light matching color, with the
  effect of shading data with significant period-5
  variations for $\Jp\lesssim\Jcb$.
  Panel (b1) shows the growth of the entanglement entropy
  $S$ vs. $1/D$ at specific cuts $x$ along the ladder.
  The cuts corresponds to specific values values of $\Jp$
  [color matched with the vertical markers in panel (a)].
  The range $\Jp<\Jca$ is well converged (blue), whereas
  the range $\Jp>\Jca$ still shows (minor) growth of the
  entanglement entropy for the largest $D$ employed [see
  remainder of data in (b), and also the increase in (a)
  from the last forward ($\gg$; blue dots) to the last
  backward sweep ($\ll$; red dots)].
  Panel (b2) shows the largest discarded DMRG weight
  $\delta\rho_\mathrm{disc}$ over the entire ladder
  vs. $1/D$, together with the convergence $\delta e_0$
  of the overall ``ground state'' energy $e_0 \equiv E_0/L$.
  Panel (b3-4) show the entanglement spectrum in the
  gapped and the intermediate phase at $\Jp=0.575$ and
  $\Jp=0.590$, respectively. The number on top
  of each level indicates the multiplet degeneracy.
  Panel (c) shows the $\SdotS$ nearest-neighbor
  correlations where the shading again indicates period-5
  variations of the corresponding data in matching color.
  A zoom into these variations around $\Jp\simeq 0.576$
  [vertical solid gray marker] is shown in the inset.
  Panel (d) shows the flow of entanglement spectra along
  the ladder, i.e. vs. $\Jp$ where different color
  represent different symmetry sectors (see legend). The
  lines connect data from every 5th bond, where bonds
  corresponding to a block size of a multiple of 5 are
  shown in strong colors.  The inset shows a semilog plot
  of the same data, yet relative to the lowest level for
  each bond.
}
\label{fig:SU5_scan}
\end{figure*}

\begin{figure*}[tbp]
\begin{center}
  \includegraphics[width=0.85\linewidth]{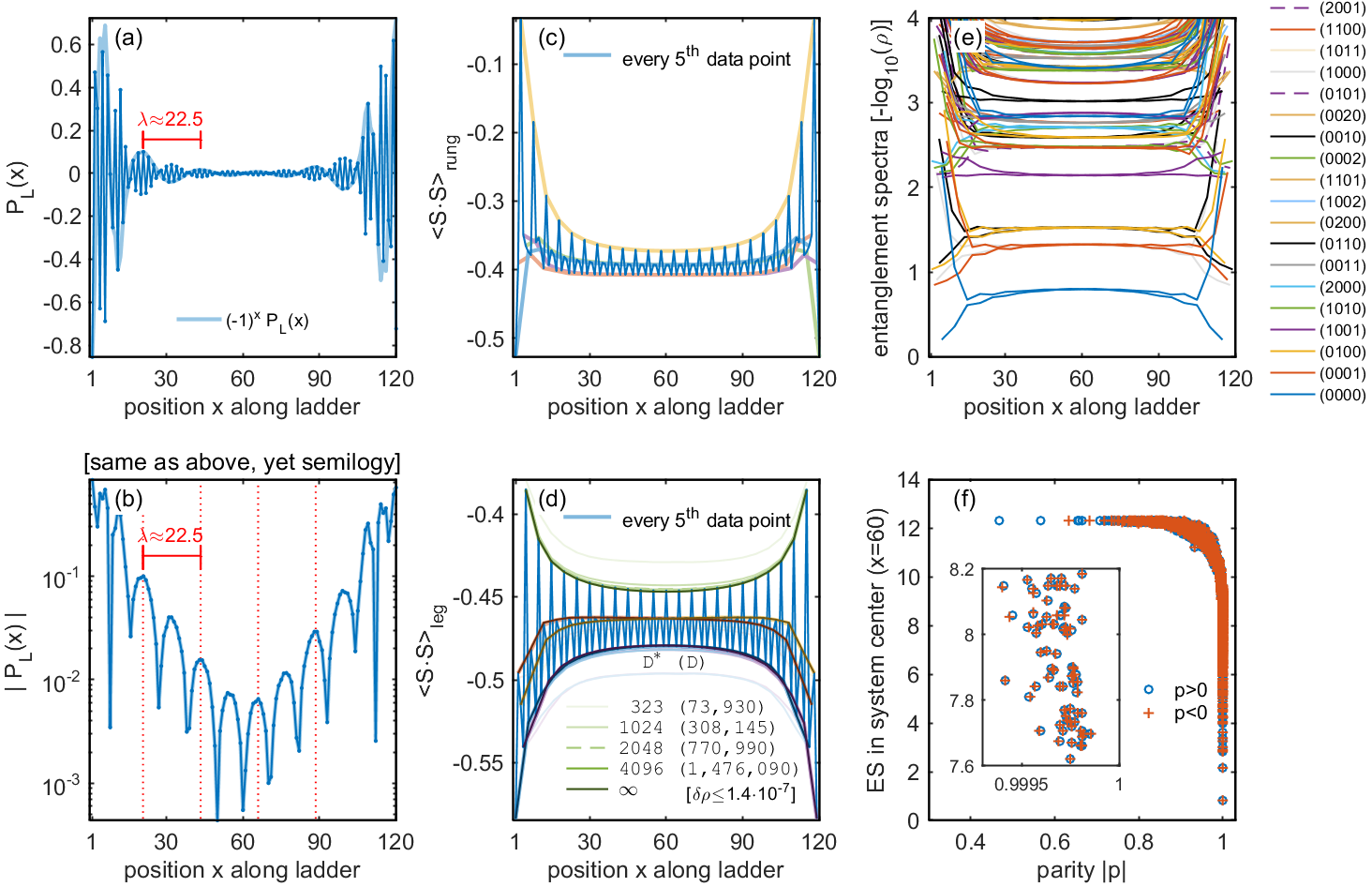}
\end{center}
\caption{
   Uniform \SU{5} ladder in the middle of the intermediate phase
   at $\Jp=0.590$. The \SU{N=5} DMRG simulation was performed
   on a ladder of length $L=120$ with open boundary condition
   keeping up to $D{^\ast}=4096$ multiplets (about $D \leq 1.5$
   million states). Panel (a) and (b) analyze the block 
   parity $P_L(x)$. Up to an alternating behavior, and within
   finite-size variations [cf. \Fig{fig:SU3_incommX}(d)],
   a period of $\lambda \sim 2N = 10$ rungs is observed. 
   This period is also marked via vertical dotted lines in (b).
   Panels (c) and (d) show the nearest-neighbor $\SdotS$
   correlation along rungs and legs, respectively.
   Panel (d) shows convergence within DMRG by adding 
   data lines for intermediate sweeps at $D^\ast$ as specified
   with the legend to the right, together with an extrapolation
   to $D^\ast \to \infty$. This demonstrates that the last
   sweep at $D^\ast=4096$ is already well converged.
   Panel (e) shows the entanglement spectra along the ladder
   where each line connects data from every bond, 
   while combining data for all ladder positions $x$
   [therefore e.g. only iterations where $x$ includes an
   integer multiple of 5 rungs support the scalar $(0000)$
   symmetry sector].  Colors themselves indicate individual
   symmetry sectors as indicated with the legend.
   Again as with \Fig{fig:SU5_scan}(d), the number $n_Y$
   of boxes in the corresponding Young tableau also identifies
   the rung position $x$ modulo $N$ along the ladder,
   i.e. $\mathrm{mod}(n_Y,N)=\mathrm{mod}(2x,N)$,
   resulting in 5 data sets w.r.t. rung position. 
   Panel (f) shows block parity vs. energy in the entanglement
   spectrum (see text) at fixed bond position $x=60$
   in the system center based on $-\log_{10}(\rho)$.
   The inset shows a zoom around
   $ES\sim 8$ which demonstrates that a systematic
   pairing of multiplets with opposite rung parity
   occurs all the way down to 
   eigenvalues of the block density matrix
   well below $10^{-8}$. 
}
\label{fig:SU5_comm}
\end{figure*}

\begin{figure*}[tbp]
\begin{center}
\includegraphics[width=0.90\linewidth]{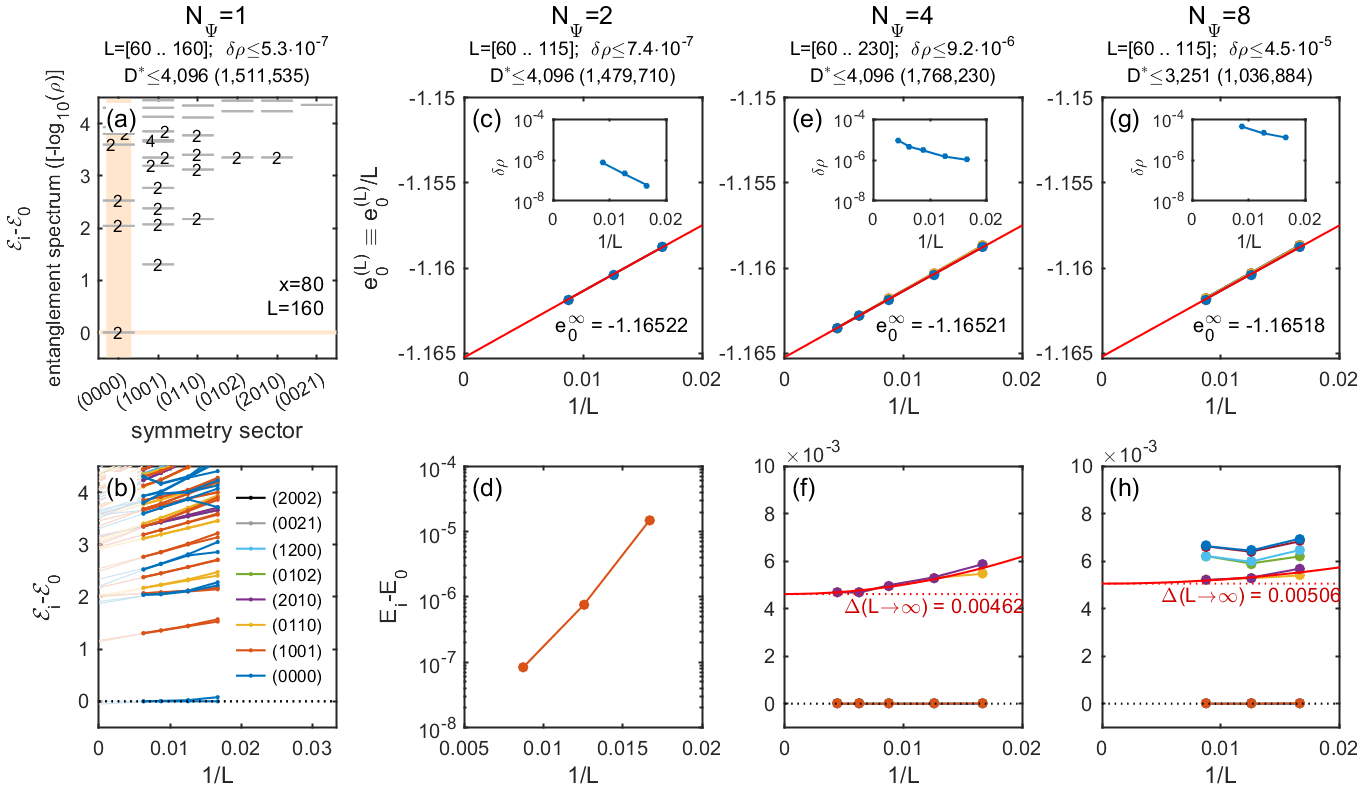}
\end{center}
\caption{
  Analysis of singlet gap for uniform \SU{5} ladders
  up to $230$ rungs at $\Jp=0.590$,
  i.e. in the middle of the intermediate phase.
  From left to right, respectively, the simulations
  targeted a total of $N_\psi=1,2,4,8$ low-energy
  states in the singlet sector $(0000)$.
  The panels in the left column, (a-b), 
  show entanglement spectra in the center of the
  ladder for the ground state only.
  In contrast to \Fig{fig:SU5_scan}(b4),
  panel (a) is derived from a uniform system here.
  The data as in panel (a) was computed for 
  system sizes up to $L=160$ and plotted vs.
  $1/L$ in panel (b). The remainder of the panels
  show DMRG results that simultaneously targeted
  multiple states. The upper panels (c-g) show
  ground state convergence,
  with the maximum discarded weight vs. $1/L$
  shown as inset to each panel. The lower
  panels (d-h) show the energy of the excited
  states relative to the ground state vs. $1/L$,
  together with a quadratic extrapolation
  towards the thermodynamic limit, $L\to\infty$.
}
\label{fig:SU5_NPsi}
\end{figure*}

The \SU{5} intermediate phase can already be observed
in a DMRG scan across the intermediate phase, as shown
in \Fig{fig:SU5_scan}. It is surrounded by a gapped
phase for $\Jp<\Jca$, and a critical phase for $\Jp>\Jcb$.
Its extremely narrow range $\Jp \in [\Jca,\Jcb] = [\jpa,
\jpb ]$ is marked by the vertical dashed lines in
\Fig{fig:SU5_scan}(a,b). In contrast to the \SU{3}
ladder, the entire \SU{5} intermediate phase ranges over a
less than 2.6\% variation in $\Jp$. The behavior of the
phase boundaries $\Jca\simeq\jpa$ and $\Jcb\simeq\jpb$ has
been checked vs. system size $L$ [cf. inset to
\Fig{fig:SU5_DMRG}(a)]. Somewhat reminiscent of the \SU{3}
ladder, the position $\Jcb$ of the upper boundary turned
out very stable, whereas the position $\Jca$ of the lower
boundary moved to slightly smaller values still with
increasing system size.  In this sense, the intermediate
phase is stable in the thermodynamic limit.

The entanglement entropy in \Fig{fig:SU5_scan}(a) shows
a sharp increase in the intermediate phase which makes
this phase numerically challenging.
As shown in the finite $D$ scaling of the entanglement
entropy in \Fig{fig:SU5_scan}(b1), the ladder
is clearly converged in the gapped regime (blue data).
In contrast, for both the intermediate as well as the 
critical large $\Jp$ regime, the DMRG simulation
is still not fully converged, despite keeping up
to $D=711,159$ states ($D^\ast=2,048$ multiplets). 
Nevertheless, in the final sweep of the DMRG simulation
the largest discarded weight over the entire chain
is already as low as $\delta\rho_\mathrm{disc}
\lesssim 10^{-6}$ [\Fig{fig:SU5_scan}(b2)].
Hence the physical picture may already be considered
converged. From (b1) one can also see that in order
to actually see the intermediate phase in the DMRG
simulation, already a large number states was
required to start with ($1/D \lesssim 3\cdot 10^{-5}$,
i.e. $D\gtrsim 30,000$).

The entanglement spectrum (ES) in the gapped VBC phase
at $\Jp<\Jca$ [\Fig{fig:SU5_scan}(b3)] features a
large gap between the lowest and first excited level
with a sparse set of higher levels, as expected.
By comparison, the entanglement spectrum of the intermediate
phase at $\Jp=0.590$ [\Fig{fig:SU5_scan}(b4)]
shows a reduced gap in the entanglement spectrum,
and a denser set of higher lying levels.
Much more remarkably, though, it shows a systematic
two-fold degeneracy in its multiplet structure.
Note that this is degeneracy of multiplets which
themselves have further large internal dimensions:
  $d(0000)=1  $,
  $d(1001)=24 $,
  $d(0110)=75 $,
  $d(0102)=126$,
  $d(0021)=175$,
etc.

\FIG{fig:SU5_scan}(c) shows the nearest-neighbor $\SdotS$
correlations along the ladder. 
For large $\Jp$, the translational symmetry breaking
along the chain diminishes [see \Fig{fig:SU5_scan}(c),
for $\Jp>\Jcb$, except close to the open right boundary].
This is consistent with a gapless, i.e. critical phase.
In the gapped VBC phase
$\Jp<\Jca$ a clear 5-rung periodicity is observed (see
inset). In particular, the $\SdotS$ bonds along the legs
(orange) clearly break up into blocks of 5 rungs, i.e.
they have 4 consecutive tight bonds followed by a weak
link (see also cartoon in \Fig{fig:overview}). This VBC
block size is natural as it represents the smallest unit
in given \SU{5} ladder that can harbor a singlet state.
In stark contrast, while also the rung data (blue)
exhibits a 5-rung periodicity, it is the {\it central}
rung within each VBC unit that is bound most weakly
(compare blue to orange data in inset), hence gathers
most of the weight from the presence of a symmetric rung
multiplet.  This is in agreement with the AASAA cartoon
picture discussed in \Sec{sec:SU3_incomm} that reduces
to an approximate resonating valence bond picture ASA
on the circumference in the case of \SU{3}
[cf.~\Fig{fig:SU3_DMRG}(a)].

The physical state changes strongly across the intermediate
phase. E.g. when zooming out of \Fig{fig:SU5_scan}(c) to
show a more extended range $\Jp \in [0,1]$, the transition in
$\SdotSrung$ almost looks like a jump to values that
differ by about a factor of 2. Eventually, however, the
intermediate phase allows for a smooth transitioning.
In contrast to the critical phase for large $\Jp$,
the symmetry breaking appears to persist along the
intermediate phase, even though it is further reduced when
analyzing larger uniform \SU{5} ladders [cf. inset to
\Fig{fig:SU5_DMRG}(a)].

\FIG{fig:SU5_scan}(d) shows the entanglement spectra along
the ladder, i.e. an ``entanglement flow diagram'' vs. $\Jp$.
The individual lines connect data from every 5th bond,
where the data from bonds the correspond a block
size of a multiple of 5 are shown in strong colors.
These entanglement spectra again also differ starkly for
the intermediate phase as compared to the small and large
$\Jp$ phases. The 2-fold multiplet degeneracy already seen
in \Fig{fig:SU5_scan}(b4) is persistent throughout the
entire intermediate phase.
The data in the main panel \Fig{fig:SU5_scan}(d) is
replotted in its inset relative to the lowest level for
each bond. This shows that the splitting of the lowest
pair of levels drops quickly to extremely small values
within the intermediate phase.
  
The \SU{5} ladder is {\it commensurate} in short-ranged
correlations for uniform ladder as, e.g.  as shown in
\Fig{fig:SU5_comm} at $\Jp=0.590$ in the middle of the
intermediate phase. In particular, the nearest-neighbor
$\SdotS$ correlations along both, rungs
[\Fig{fig:SU5_comm}(c)] and legs [\Fig{fig:SU5_comm}(d)]
show a clean 5-rung periodicity.

However, there are traces of incommensuration also in
the \SU{5} intermediate phase.  For example, the minor
oscillatory behavior towards the upper boundary $\Jcb$ in
\Fig{fig:SU5_scan}(d) [in particular, see also inset] may
be interpreted as the onset of incommensurate behavior.
This is even more pronounced still in the block parity
$P_L(x)$ shown in \Fig{fig:SU5_comm}(a) that shows a
tunable wavelength with $\Jp$. In particular, the
wavelength seen in \Fig{fig:SU5_comm}(a) becomes
shorter as $\Jp$ is reduced and moves closer to $\Jca$.
In stark contrast to the \SU{3} ladder [e.g. see
\Fig{fig:SU3_incommX}(d)], the data in \Fig{fig:SU5_comm}(a)
decays approximately exponentially towards the bulk of the
ladder, as seen in the semilog plot in
\Fig{fig:SU5_comm}(b).
However, an analysis along the lines of
\Fig{fig:SU3_incommX} for the \SU{3} ladder earlier did
not support a global signature of incommensurate behavior for
the \SU{5} ladder (data not shown). Specifically, there
was no indication for an alternation in the ground state
parity with increasing system size.

The rapid systematic pairing of multiplets away from the
open boundary was already seen in the entanglement flow in
the DMRG scan in \Fig{fig:SU5_scan}(d), and is again also
encountered for uniform ladders as in \Fig{fig:SU5_comm}(e).
The origin of this systematic pairing in the intermediate
phase is related to the rung parity, in that it is multiplets
of opposite rung parity that pair up. This is demonstrated in
\Fig{fig:SU5_comm}(f). The systematic pairing and correlation
with the rung parity holds for states with weight down to
less than $10^{-8}$ in the reduced density matrix!
For \Fig{fig:SU5_comm}(f),
the block density matrix $\rho(x)$ for
sites $x'>x$ for fixed bond $x=60$ in the system center
was compared to the matrix elements $P(x)$ of the swap
operator for all rungs $x'<x$ in the DMRG basis
at bond $x$ [therefore, e.g. $P_L(x)=\trace{\rho(x) P(x)}$].
The simultaneous (approximate) diagonalization of both,
$\rho(x)$ and $P(x)$, for the data in \Fig{fig:SU5_comm}(f)
was performed separately for each individual
symmetry sector as follows:
(i) the effective many body state space at bond $x$ was
split into symmetric ($p>0$) and antisymmetric ($p<0$)
states by first diagonalizing $P(x)$.
(ii) Then the reduced density matrix $\rho(x)$ was mapped
into these parity sectors, and diagonalized to obtain the
entanglement spectra [ES, using $-\log_{10}
\mathrm{eig}(\rho)$; vertical data in \Fig{fig:SU5_comm}(f)],
as well as the final basis. The latter
(iii) was used to compute the expectation values of the
parity $p$ [horizontal data in \Fig{fig:SU5_comm}(f)]
which via (i) results in values very close to $\pm 1$
paired up in energy.

In order to tackle the question whether the intermediate
phase is gapped, we compute excited states within our
DMRG, followed by finite size scaling.  The results are
shown in in \Fig{fig:SU5_NPsi}.  There we compare data
from DMRG simulations in the intermediate phase at
$\Jp=0.590$ that simultaneously targeted $1,2,4$, and
$8$ states from left to right panels, respectively.
All states belonged to the singlet sector, therefore the
following discussion concerns the singlet gap of the
\SU{5} ladder. The scaling of the entanglement spectra
in the center of ground states for ladders of length $L$
in \Fig{fig:SU5_NPsi}(b) shows a pronounced entanglement
gap that persist in the thermodynamic limit when
extrapolating vs. $1/L\to 0$.

When simultaneously targeting multiple state, the matrix
product state (MPS) of DMRG \cite{Affleck88, Rommer95,
Verstraete04_vbs, Schollwoeck05, Schollwoeck11} acquires
an additional index that references the individual states.
Depending whether this index is associated with the left
or the right of a given bond in the ladder, the
entanglement spectra looks different.  Hence when
targeting $N_\Psi>1$ states, entanglement spectra become
less useful.  Therefore the remainder of the panels,
\Fig{fig:SU5_NPsi}(c-h), focus on the analysis of global
energy convergence.
The upper panels, \Fig{fig:SU5_NPsi}(c,e,g), show a
consistent convergence of the ground state energy per
site, suggesting $e_0^{\infty} = -1.16522$ [most accurate
value for $N_\Psi=2$, panel (c); larger $N_\Psi$ for the
same or smaller bond dimension $D$ naturally leads to
slightly less accurate values, as also seen in the overall
trend with increasing $N_\Psi$ in the discarded weights in
the insets].

The lower panels, \Fig{fig:SU5_NPsi}(d,f,h) analyze the
energy of the excited energy eigenstates relative to the
ground state energy.  The systematic 2-fold  degeneracy of
the entanglement spectra for the ground state [cf.
\Fig{fig:SU5_NPsi}(a)] is inherited when computing excited
energy eigenstates of the full system.  This is plausible
given the interpretation obtained from
\mbox{\Fig{fig:SU5_comm}(e-f)} that the systematic
degeneracy of the entanglement spectra is related to
states of opposite rung-parity.  Hence the energy
splitting between ground state and first excited state
diminishes in the thermodynamic limit, as seen in
\Fig{fig:SU5_NPsi}(d).

When including the next two degenerate states, the finite
size scaling in \Fig{fig:SU5_NPsi}(f) reveals that this
state remain separated by a small but finite energy gap of
about $\Delta\simeq 0.004$ in the thermodynamic limit.
When including another four states in the last column of
\Fig{fig:SU5_NPsi}(g-h), these additional states are
already close to the excited states encountered in
\Fig{fig:SU5_NPsi}(f).  Therefore even though the
estimated gap $\Delta$ is on the order of $1/L$ for the
largest system size in \Fig{fig:SU5_NPsi}(f), we interpret
the weak dependence of the energy splitting on $1/L$,
indeed, as indication for a small but finite energy gap.

\section{DMRG study of the $N=3$ and $N=5$ large $\Jp$ regime}
\label{app:largeJp}

\begin{figure}[tb!]
\begin{center}
  \includegraphics[width=0.9\linewidth]{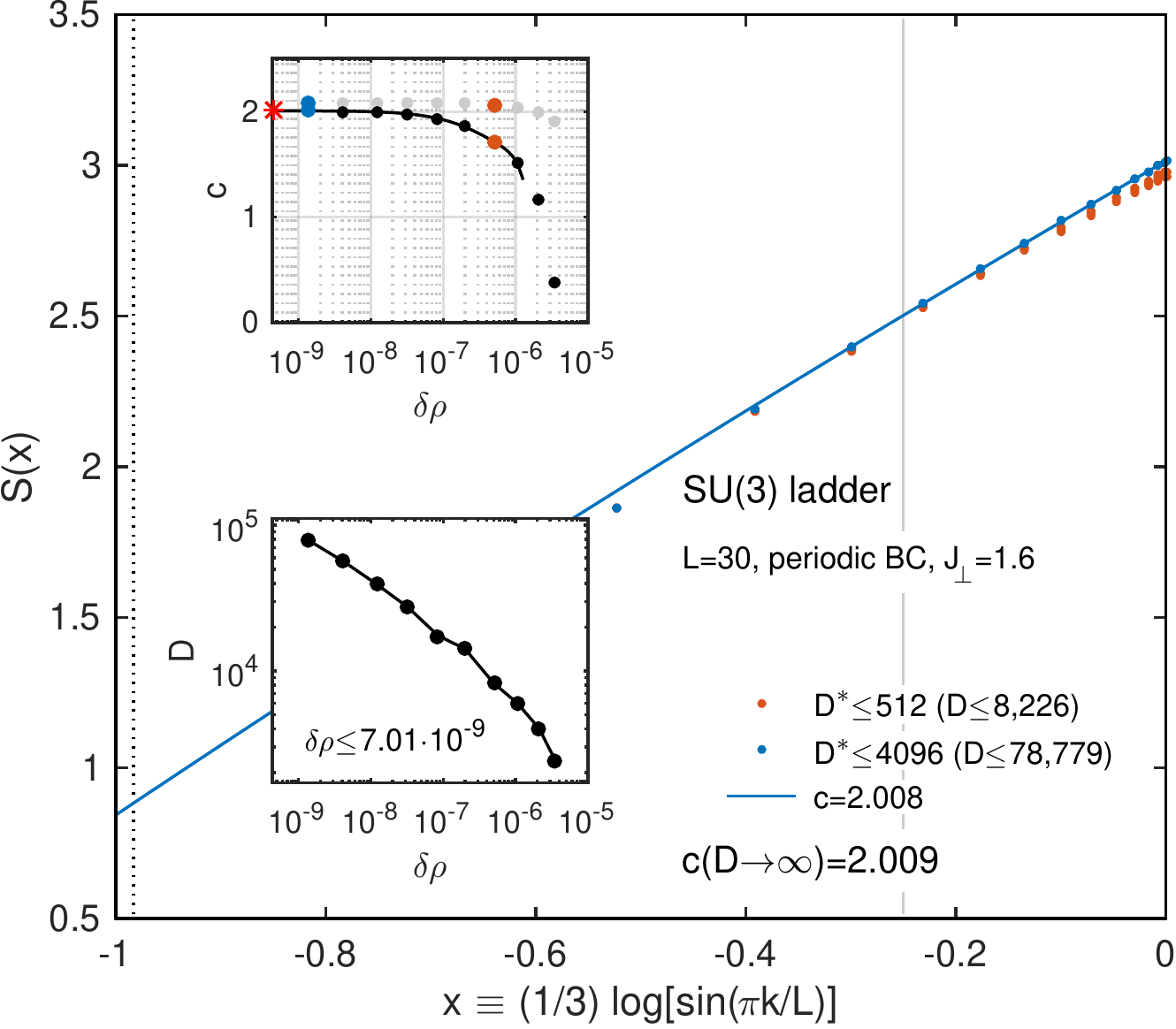}
\end{center}
\caption{
  Central charge for \SU{3} ladder in the large $\Jp$ phase,
  actually having $\Jp=1.6$ just above $\Jcb=1.562$,
  based on the scaling of the block entanglement entropy
  using periodic boundary conditions
  (see \cite{Calabrese09,Amico08} and also text).
  For fixed number of $D\to\infty$ kept states
  ($D^\ast$ multiplets), quadratic fit
  for the range $x>x_0=-0.25$ (see vertical gray
  marker) e.g. results in the blue line for the
  last DMRG sweep performed. Its slope
  at $x=0$ yields the central charge
  $c=2.008$ as indicated with the legend.
  The left dotted marker indicates $(1/3)\log(L/\pi)$.
  Therefore with $4$ data points between
  this marker and the one at $x_0$, the overwhelming
  part of the ladder ($22$ rungs out of $30$ rungs)
  fall into $x>x_0$).
  The upper inset analyses the convergence of the central
  charge vs. average discarded weight $\delta\rho$
  for given DMRG sweep at fixed $D$. The data was
  obtained by evaluating the slope of $S(x)$ as shown
  in the main panel for fixed $D$ at $x=x_0$ and $x=0$
  (gray and black data, respectively).
  The red asterisk corresponds to the extrapolated
  value $D\to\infty$ as specified with the legend
  in the main panel.
  The lower inset shows discarded weight $\delta\rho$
  vs. number $D$ of kept states.
}
\label{fig:SU3_largeJp_c}
\end{figure}


\begin{figure}[tb!]
\begin{center}
  \includegraphics[width=0.9\linewidth]{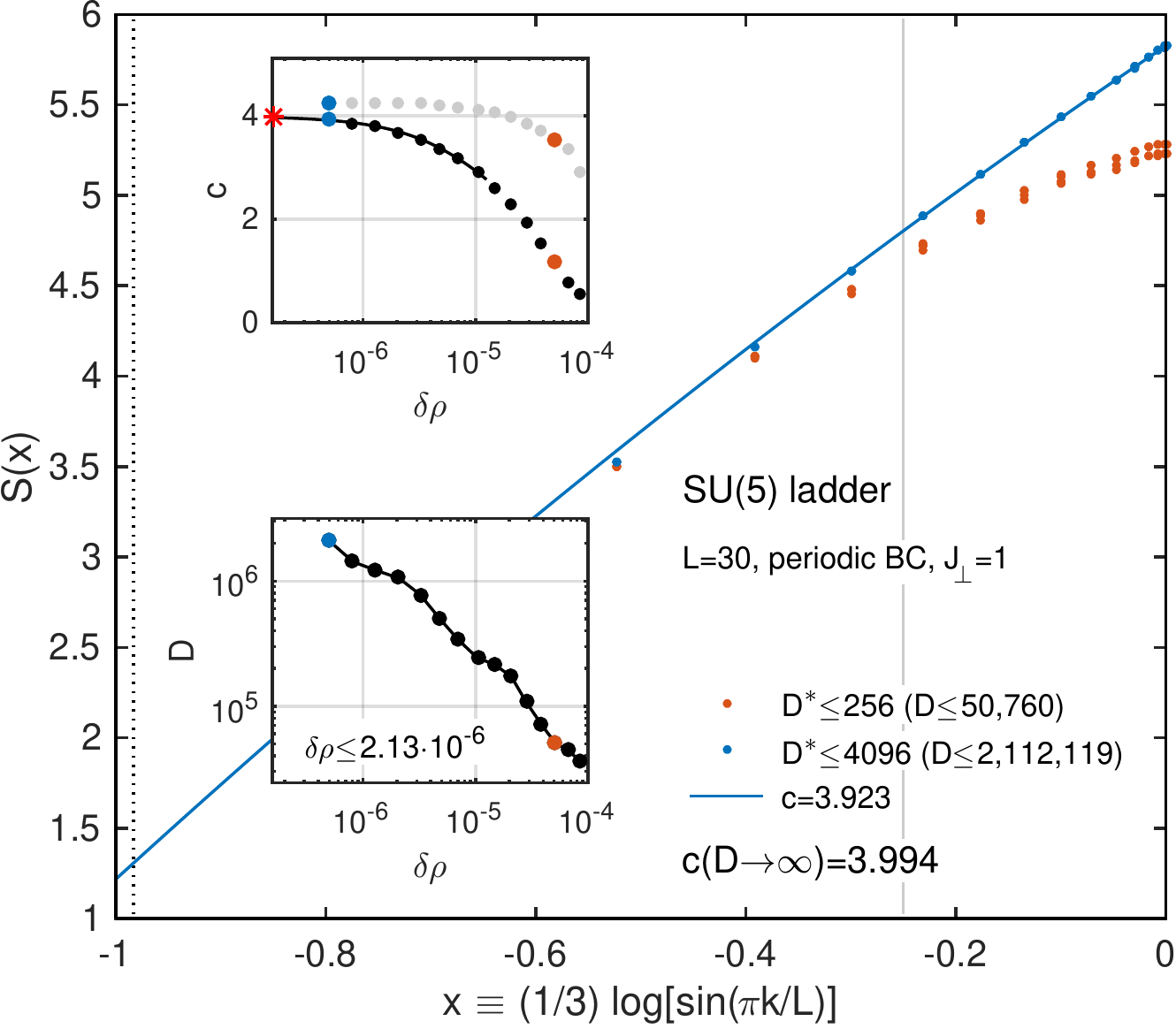}
\end{center}
\caption{
  Central charge for \SU{5} ladder in the large $\Jp$
  phase for $L=30$, with an analysis of the entanglement
  entropy that is completely
  analogous to the case of \SU{3} in \Fig{fig:SU3_largeJp_c}.
  In the extrapolated limit $D\to\infty$ and by evaluating
  the slope at $x=0$, this yields a central charge of
  $c=3.994$ (see legend and also red asterisk in upper inset),
  in excellent agreement with the expected value of $c=4$.
}
\label{fig:SU5_largeJp_c}
\end{figure}


For large rung coupling $\Jp$, as explained in the main
text, an \SU{N} ladder maps onto an effective chain where
each site transforms according to the antisymmetric irrep
(\,${\tiny \yng(1,1)}$\,) of \SU{N}. For odd $N$ (such as
$N=3$ and $N=5$), such a system is predicted to be
critical and can be described by conformal field theories.
Numerically,  the corresponding central charges can be
determined from the scaling of the block entanglement
entropy of a contiguous block of length $k$ rungs for a
ladder of length $L$ with periodic boundaries,
\cite{Calabrese09,Amico08},
\begin{eqnarray}
   S(k) &=& \tfrac{c}{3}
      \log\bigl[ \tfrac{L}{\pi}\sin(\tfrac{\pi k}{L})\bigr]
   + \mathrm{const}  \\ &\equiv&
    c \cdot \underbrace{\tfrac{1}{3}
   \log\bigl[ \sin(\tfrac{\pi k}{L})\bigr]}_{\equiv x}
   + \mathrm{const}'
\text{.}\label{eq:ccharge}
\end{eqnarray}
This allows to extract the central charge $c$ from the
slope of $S(x)$ around $x\simeq 0$, which is equivalent to
evaluating the curvature of $S(k$) in the center of the
system.  Having periodic boundary conditions, finite size
effects are strongly reduced as compared to open boundary
conditions, hence fairly short systems often suffice (here
$L=30$). This is important, since the entanglement entropy
is about twice as large as compared to a system with open
boundary conditions. In the DMRG simulations, in practice,
the periodic boundary was dealt with by folding a circle
of $L$ rungs into a double ladder of length $L/2$ each,
that are connected at the ends and otherwise have
interleaved rungs from the upper and lower half-ladder,
respectively. This construction avoids a long range
coupling term in the Hamiltonian across the entire system.
The resulting system then, while expensive in any case,
has effectively an open left and right boundary 
with at most 4th-nearest-neighbor interactions
which is beneficial for DMRG.

The central charge is evaluated for subsequent DMRG sweeps
at fixed $D$ each, where $D$ is ramped exponentially $D
\to \sqrt{2}D$ from one sweep to the next. Each sweep also
results in some average discarded weight $\delta\rho$.
Since the behavior of $\delta\rho$ vs.~$D$ is somewhat
irregular [see lower inset in \Fig{fig:SU3_largeJp_c}
and, in particular, also later for the case of \SU{5}
in \Fig{fig:SU5_largeJp_c}], the analysis of the central
charge is performed vs. $\delta\rho \to 0$ and not vs.
$1/D\to 0$. This results in much smoother data that is
better suited for quadratic polynomial extrapolation.

The resulting data for the central charge for the \SU{3}
ladder at strong coupling is shown in \Fig{fig:SU3_largeJp_c}.
Even though $\Jp$ is just across the boundary of the
intermediate phase ($\Jp=1.60 > \Jcb=1.562$), a length
$L=30$ already yields a central charge $c=2$ to better
than 1\%, which is in agreement with the expected
central charge from conformal field theory (CFT), $c=N-1$
[see \Fig{fig:overview} in the main text]. From the
analysis in the upper inset (gray vs. black data) one can
conclude, that in given case a periodic ladder of up to
$L=9$ rungs, i.e. significantly shorter than the chosen
$L=30$, already also would yield a central charge
reasonably close to $c\simeq 2$. But uncertainty margins
are significantly reduced by going to somewhat larger
system sizes.

Similarly, the \SU{5} ladder at large rung coupling $\Jp$
becomes critical with a central charge $c=N-1=4$.  By
analysis of the entanglement entropies, this is confirmed
in \Fig{fig:SU5_largeJp_c} at $\Jp=1$, again to within an
uncertainty of less than 1\%. This calculation explicitly
goes beyond a million of kept states in the DMRG.
Once all symmetry related generalized Clebsch-Gordan
coefficients are computed once and for all
\cite{Wb12_SUN}, this calculation runs an a regular
state-of-the-art workstation within several days.

\end{document}